\documentclass[journal]{IEEEtran}
\usepackage{graphicx}
\usepackage{bm}
\usepackage{cite}
\usepackage{pgfplots}
\usepackage{color}
\usepackage{algorithm}

\usepackage{amsmath,amssymb,amsthm,dsfont,mathrsfs}

\def\diag{{\mathrm{diag}}}
\def\tr{{\mathrm{tr}}}
\def\A{{\mathbf{A}}}
\def\Y{{\mathbf{Y}}}

\def\I{{\mathbf{I}}}

\def\B{{\mathbf{B}}}

\def\y{{\mathbf{y}}}

\def\a{{\mathbf{a}}}
\def\n{{\mathbf{n}}}

\def\X{{\mathbf{X}}}
\def\Y{{\mathbf{Y}}}
\def\h{{\mathbf{h}}}

\def\c{{\mathbf{c}}}

\def\w{{\mathbf{w}}}
\def\W{{\mathbf{W}}}

\def\Z{{\mathbf{Z}}}
\def\z{{\mathbf{z}}}

\begin{document}

\title{Joint Channel Estimation and User Grouping for Massive MIMO Systems}

\author{
        Jisheng~Dai, An~Liu, and Vincent K. N.  Lau


\thanks{J. Dai is with the Department of Electronic Engineering, Jiangsu University, Zhenjiang 212013, China, and was with the Department of Electronic and Computer Engineering, Hong Kong University of Science and Technology, Hong Kong (e-mail: jsdai@ujs.edu.cn).}
\thanks{A. Liu is with the College of Information Science and Electronic Engineering, Zhejiang University, Hangzhou 310027, China (e-mail: anliu@zju.edu.cn).}
\thanks{V. K. N. Lau is with the Department of Electronic and Computer Engineering, Hong Kong University of Science and Technology, Hong Kong (e-mail: eeknlau@ust.hk).}
}

\maketitle

\begin{abstract}
This paper addresses the problem of joint downlink channel estimation and user grouping in  massive multiple-input multiple-output (MIMO) systems, where the motivation comes from the fact that the channel estimation performance can be improved if we exploit additional common sparsity among nearby users.
In the literature, a commonly used group sparsity model assumes that users in each group share
a uniform sparsity pattern. In practice, however, this oversimplified assumption usually fails to hold, even
for physically close users.
Outliers deviated from the uniform sparsity pattern in each group may
significantly degrade the effectiveness of common sparsity, and hence bring limited (or negative) gain for channel
estimation.
To better capture the group sparse structure in practice, we provide a general model having two sparsity components: commonly shared sparsity and individual
sparsity, where the additional individual sparsity accounts for any outliers. Then, we propose a novel sparse Bayesian learning (SBL)-based framework
to address the joint channel estimation and user grouping problem under the general sparsity model. The framework  can fully exploit the common sparsity among nearby users and
exclude the harmful effect from outliers simultaneously. Simulation results reveal substantial performance gains over
the existing state-of-the-art baselines.

\end{abstract}


\IEEEpeerreviewmaketitle

\section{Introduction}

Massive multiple-input multiple-output (MIMO) can support high spectrum and energy efficiency, and it has been widely considered as
one of the key candidate technologies to meet the capacity demand for the next generation of wireless communications \cite{marzetta2010noncooperative,larsson2014massive, rusek2013scaling}.
To fully harvest the benefit of excessive base station (BS) antennas,  knowledge of channel state information at the transmitter
(CSIT) is an essential requirement  \cite{lu2014overview, shen2017high}.
However, it is challenging to acquire the accurate CSIT, since the training overhead for CSIT acquisition grows proportionally with the number of BS antennas, which can be very large in such systems.
Early works sidestep this challenge
by adopting a time-division duplexing (TDD) model \cite{lu2014overview,hoydis2013massive}, where the CSIT can be obtained by exploiting channel reciprocity, and
the uplink pilot-aided training overhead is only proportional to the number of mobile users.

Unfortunately, channel reciprocity does not hold for massive MIMO systems with a frequency-division duplexing (FDD) model. Compared with a TDD system, an FDD system has its own advantages; e.g., it can provide more efficient communications with low latency \cite{xu2014user,gao2016structured}. FDD also dominates current cellular systems, and for 5G wireless networks, the radio capability for FDD remains in the specifications \cite{3gpp-5G}.
Therefore, it is also important to consider CSIT acquisition for FDD systems.
In fact, there are situations when it is necessary to consider downlink channel estimation even in TDD systems. For example, due to random radio-frequency (RF) circuit mismatches in the uplink and downlink and limited coherence time, the channel reciprocity may no longer hold \cite{mi2017massive,bjornson2014massive}. Moreover, in LTE/5G systems, there exist situations when users only use some of the antennas to transmit in the uplink. In this case, the channel associated with the other user antennas has to be estimated via downlink training. In addition, the cell-edge users may suffer from very low SNR in the channel estimation phase due to the limited power budget at the mobile device. In this case, it is preferable to use downlink channel estimation because the BS can transmit pilot signals at a larger power to meet the channel estimation accuracy.

Many works have shown that the effective dimension of a massive MIMO channel is actually much less than its original dimension because of the limited local scattering effect in the propagation environment \cite{gao2015spatially,rao2014distributed,hoydis2012channel,liu2017closed}. Specifically,  the massive MIMO channel has an approximately sparse representation under the discrete Fourier transform (DFT) basis if the BS is equipped with a large uniform linear array (ULA) \cite{wen2015channel, rao2014distributed,chen2016pilot,shen2016compressed}. As a consequence, a large number of compressive sensing (CS) algorithms that exploit the hidden sparsity under the DFT basis have been proposed for downlink channel estimation and feedback \cite{gao2015spatially,rao2015compressive,rao2014distributed,choi2014downlink,you2016channel,gao2016structured,gao2015priori,liu2016exploiting,liu2017closed}. Nevertheless, there are at least two challenges of the DFT-based methods:
1) they are only applicable to ULAs because the sparse property hinges strongly on the shared structure between the DFT basis and the ULA steering;
and 2) they always suffer from inevitable modeling error caused by direction mismatch.
To alleviate the  modeling error, a denser sampling grid covering the angular domain with more points (named overcomplete DFT basis) was considered in \cite{ding2015channel,ding2015compressed,ding2016dictionary}. However, the overcomplete DFT method is still applicable to ULAs only, and it may lead to a high performance loss if the grid is not sufficiently dense.

Recently, the sparse Bayesian learning (SBL) method has attracted significant attention for sparse signal recovery \cite{wipf2004sparse,tipping2001sparse,dai2018sparse,ji2008bayesian,dai2018fdd,yang2013off}. The SBL-based framework has an inherent learning capability,
and hence, no prior knowledge about the sparsity level, noise variance or direction mismatch is required.  Moreover, theoretical and empirical results have shown that SBL methods can achieve better performance than the $l_1$-norm-based methods \cite{wipf2004sparse,ji2008bayesian}.
Our previous work \cite{dai2018fdd} introduced an off-grid SBL-based method for downlink channel estimation, which can be applied to arbitrary 2D-array geometry and substantially reduces the modeling error caused by direction mismatch.
The method in \cite{dai2018fdd} overcomes all the aforementioned challenges of the DFT-based methods, and simulation results
illustrated that it can achieve much better channel estimation performance than the existing state-of-the-art methods.
However, \cite{dai2018fdd} only focused on single-user channel estimation in massive MIMO systems.

Many studies have observed that channels of multi-user massive MIMO systems
may share  common sparsity structures due to the commonly shared local scattering clusters \cite{hoydis2012channel,gao2011linear}.
To exploit the common sparsity among nearby users,
a joint orthogonal matching pursuit recovery algorithm was proposed in \cite{rao2014distributed}.
However, the effectiveness of that approach relies on appropriate user clustering in the multi-user MIMO network.
While there are various user clustering methods \cite{xu2014user,adhikary2013joint,nam2014joint} in the literature, they are targeted for different purposes. It is also worth noting that the meaning of group sparsity from the perspective of compressed sensing (CS) \cite{simon2013sparse,baraniuk2010model} is different from the one used in this paper. In CS, group sparsity is usually known as block
sparsity, which means the locations of significant coefficients cluster in blocks under a known specific sorting order.
To the best of our knowledge, user clustering for maximizing the common sparsity has not been investigated before.
In this paper,  we propose an efficient off-grid SBL-based approach for joint channel estimation and user grouping
to enhance the effectiveness of common sparsity in massive MIMO systems. The following summarizes the contributions of this paper.

\begin{itemize}

  \item {\bf General Sparsity Model for User Grouping}

  We develop a more  general  sparsity model to better capture the group sparse structure  in practical multi-user massive MIMO systems. In the literature, a commonly used group sparsity model assumes that users in each group share a uniform sparsity pattern \cite{wang2016novel}.
  This oversimplified model can simplify the procedure for user grouping; however, it usually fails to hold, even for physically close users, in practice.
  Outliers deviated from the uniform sparsity pattern in each group may significantly degrade the effectiveness of the common sparsity, and bring  limited (or negative) gain for channel estimation.
  To address this issue, we propose a general model having two sparsity components: a commonly shared sparsity and an
  individual sparsity. Since the additional individual sparsity can account for any outliers, the new model may capture a more complex and realistic group sparse structure in real-world applications (see Fig.~\ref{figMIMO} for example).

  \item {\bf SBL-based Framework for Joint Channel Estimation and User Grouping}

  We propose a novel SBL-based method to autonomously partition users into groups during the channel estimation under the general sparsity model.
  SBL-based methods have been widely applied to estimate the sparse channel in single-user massive MIMO systems, but they are not applicable to joint user grouping and channel estimation in multi-user massive MIMO systems.
  To the best of our knowledge, the method proposed for wideband direction-of-arrival estimation in \cite{wang2016novel} is the only candidate that may be tailored to solve the problem of joint channel estimation and user grouping. However, it requires the aforementioned restrictive assumption that users in each group share a uniform sparsity structure. To handle the more practical general sparsity model,
  we propose a novel SBL-based framework,
  which can fully exploit the common sparsity among nearby users and exclude the harmful effect from outliers simultaneously.
  Moreover, the grid-refining procedure used in \cite{dai2018fdd} is also extended to the framework to efficiently combat direction mismatch with an arbitrary 2D-array geometry.

\end{itemize}

The rest of the paper is organized as follows. In Section II, we present the system model and the general sparsity model.
In Section III, we provide the SBL-based method for joint channel estimation and user grouping. In Section IV,  we extend the proposed method to handling direction mismatch with an arbitrary 2D-array geometry.
Numerical experiments and a conclusion  follow in Sections V and VI, respectively.

$Notations:$ $\mathbb{C}$ denotes complex number, $\|\cdot\|_p$ denotes $p$-norm, $(\cdot)^T$ denotes transpose, $(\cdot)^H$ denotes Hermitian transpose,
$(\cdot)^\dag$ denotes pseudoinverse,
$\I$ denotes identity matrix,
$\A_\Omega$ denotes the sub-matrix formed by collecting the columns from $\Omega$,
$\mathcal{CN}(\cdot| \bm\mu, \bm\Sigma)$ denotes complex Gaussian distribution with mean $\bm\mu$ and variance $\bm\Sigma$, $\mathrm{supp}(\cdot)$ denotes the set of indices of nonzero elements, $\tr(\cdot)$ denotes trace operator, $\diag(\cdot)$  denotes diagonal operator, and $\mathrm{Re}(\cdot)$ denotes real part operator.

\section{Data Model}

\subsection{Massive MIMO Channel Model}

\begin{figure}
\center
\includegraphics[scale=0.28]{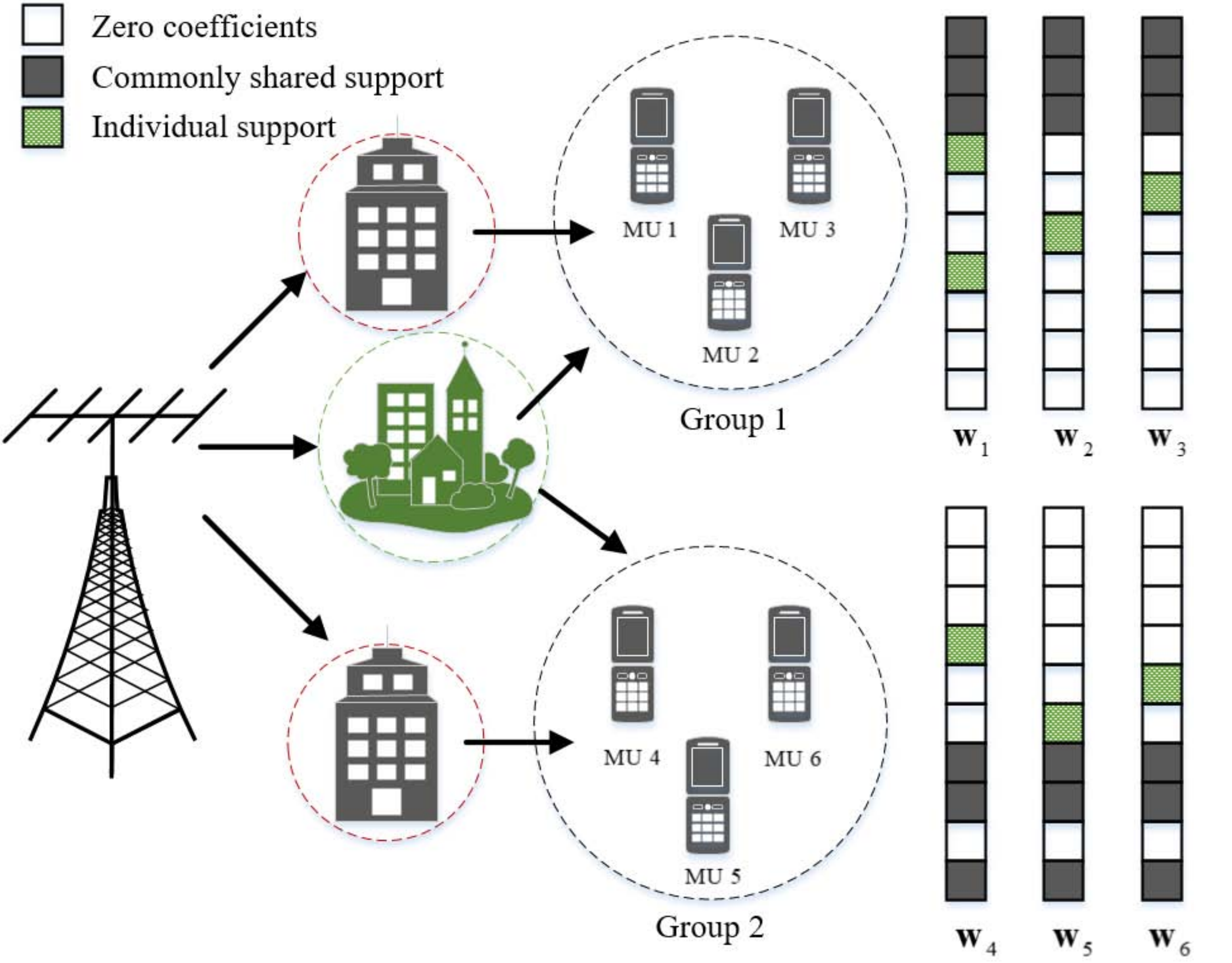}
\caption{Illustration of system model and general sparsity mode,  where the commonly shared support sets for two groups are $\mathcal{S}_1=\{1,2,3\}$ and $\mathcal{S}_2=\{7,8,10\}$, respectively.}\label{figMIMO}
\end{figure}

Consider a massive MIMO system  as illustrated in Fig.~\ref{figMIMO}.  There is one BS with $N$ $(\gg 1)$ antennas and $K$ mobile users (MUs) with  a single antenna. Assume that the BS  broadcasts a sequence of $T$ training pilot symbols, denoted by $\X\in  \mathbb{C}^{T\times N}$, for each MU to estimate the downlink channel. Then, the downlink received signal $\y_k\in \mathbb{C}^{T\times 1}$ at the $k$-th MU is given by
\begin{align}\label{eq-dly}
\y_k&= \X \h_k + \n_k,  
\end{align}
where $\h_k\in \mathbb{C}^{N\times 1} $ stands for the downlink channel vector from the BS to the $k$-th MU, $\n_k\in \mathbb{C}^{T\times 1} $ stands for the additive complex Gaussian noise with each element being zero mean and variance $\sigma^2$ in
the downlink, and $\mathrm{tr}(\X\X^H)=PTN$, with $P/\sigma^2$ measuring the training signal-to-noise ratio (SNR).
If the BS is equipped with a linear array, $\h_k$ can be formulated as \cite{tse2005fundamentals,3gpp,molisch2003geometry}
\begin{align}
\h_k&= \sum_{c=1}^{N_c} \sum_{s=1}^{N_s} \xi_{c,s}^k \a(\theta_{c,s}^k),\label{eqmo1}
\end{align}
where $N_c$ stands for the number of scattering clusters, $N_s$ stands for the number of sub-paths per scattering cluster, $\xi_{c,s}^k $  is the complex gain of the $s$-th sub-path in the $c$-th scattering cluster for the $k$-th MU, and $\theta_{c,s}^k$ is the corresponding  azimuth angle-of-departure (AoD).
For a linear array, the steering vector $\a(\theta)\in  \mathbb{C}^{N\times 1} $ is in the form of
\begin{align}
\a(\theta) &=[1,  e^{-j2\pi  \frac{d_2}{\lambda}\sin(\theta) }, \ldots, e^{-j2\pi  \frac{d_N }{\lambda}\sin(\theta)} ]^T,\label{eqabdod1}
\end{align}
where $\lambda$ is the wavelength of the downlink propagation, and $d_n$ stands for the distance between the $n$-th antenna and the first antenna. For a ULA,  $\a(\theta)$ can be simplified by
\begin{align}
\a(\theta) &=[1,  e^{-j2\pi \frac{d}{\lambda} \sin(\theta)}, \ldots, e^{-j2\pi\frac{(N-1)d}{\lambda} \sin(\theta)} ]^T,\label{eqdod1}
\end{align}
where $d$ stands for the distance between adjacent sensors.

For ease of notation, we denote the true AoDs for MU~$k$  as $\{\theta_l^k,l=1,2,\ldots, L\}$ with $L=N_cN_s$. Let $\hat{\bm\vartheta}=\{\hat{\vartheta}_{l}\}_{l=1}^{\hat{L}}$ be a fixed sampling grid that uniformly covers the angular domain $[-\frac{\pi}{2}, \frac{\pi}{2}]$, where $\hat L$ denotes the number of grid points.
If the grid is fine enough, such that all the true AoDs  $\theta^k_l$s, $l=1,2,\ldots, L$, lie on (or practically close to) the grid, we have\footnote{The DFT basis becomes a special case of $\A$ if the BS is equipped with a ULA and there are $N$ grid points such that $\{\sin\hat{\vartheta}_{l}\}_{l=1}^{\hat{L}}$ uniformly covers the range $[-1,1]$.}
\begin{align}\label{Hmodeln1}
\h_k= \A \w_k,
\end{align}
where $\A= \begin{bmatrix} \a ( \hat{\vartheta}_{1} ), & \a(\hat{\vartheta}_{2}), & \ldots, & \a (\hat{\vartheta}_{\hat L})
\end{bmatrix} \in \mathbb{C}^{N\times \hat L}  $,
and $\w_k \in \mathbb{C}^{\hat{L}\times 1} $ is a vector with a few non-zero elements corresponding to the true directions at $\{\theta_l,l=1,2,\ldots, L\}$. With (\ref{eq-dly}) and (\ref{Hmodeln1}), $\y_k$ can be rewritten by
\begin{align}\label{dlofmodel}
\y_k= \X \A \w_k + \n_k = \bm\Phi\w_k + \n_k,
\end{align}
where $\bm\Phi\triangleq\X \A$.
Note that the assumption that all true AoDs are located on the predefined spatial grid is not always
valid in practice \cite{yang2013off,dai2017root}. We will address the direction mismatch in Section IV, as well as the  extension for arbitrary 2D-array geometry.

\subsection{General Sparsity Model}

The massive MIMO channel usually has the following two important properties:
\begin{itemize}
  \item (\emph{Sparsity Property}): Due to the limited local scattering effect in the propagation environment \cite{chen2016pilot,gao2015spatially,shen2016compressed},
 the number of scattering clusters $N_c$ is usually small and the sub-paths associated with each scattering cluster are likely to concentrate in a small range. In other words, only a few angles are occupied in the angular domain, which, in return, brings a sparse representation $\w_k$.

  \item (\emph{Group  Property}): Some users may face a very similar scattering structure if they are physically close to each other \cite{hoydis2012channel,gao2011linear,rao2014distributed}. Hence, the performance of the downlink channel estimation can be improved if we can exploit the common sparsity among nearby users, because it can bring additional useful information for sparse signal recovery algorithms.

\end{itemize}

Without loss of generality, assume that the $K$ users can be partitioned into $G$ groups $\{\mathcal{G}_1, \mathcal{G}_2,\ldots,  \mathcal{G}_G \}$. The commonly used group sparsity model assumes that users in each group share a uniform sparsity pattern \cite{wang2016novel}, i.e.,
\begin{align}\label{equnifc}
\mathrm{supp}(\w_{p}) = \mathrm{supp}(\w_{q})  , ~~~~ p,q\in \mathcal{G}_g.
\end{align}
This assumption  can simplify the procedure for user grouping, but usually fails to hold in practice, because it is a restrictive constraint requiring the same scattering structure for users in each group. The channel estimation performance may be significantly degraded by the outliers deviated
from the uniform sparsity pattern in each group.

To capture a more complex and realistic group sparse structure, we partition the sparse representation vector into two parts, i.e.,
\begin{align}\label{eq-modpss}
\w_k=\w_k^s + \w_k^v,
\end{align}
where $\w_k^s$ stands for the commonly shared sparse representation vector whose support corresponds to the commonly shared support, and $\w_k^v$ stands for the individual sparse representation vector whose support corresponds to the individual support.


\noindent\textbf{Definition~1.} (\emph{General Sparsity Model}): Let the sparse representation vector  be formulated as in (\ref{eq-modpss}), and there be multiple groups, with each group having a distinct commonly shared sparsity pattern; i.e.,
  \begin{align}\label{eq-parss22}
    &\mathcal{S}_g= \mathrm{supp}(\w^s_{p})=\mathrm{supp}(\w^s_{q}), ~~\forall p,q\in \mathcal{G}_g,\\
    &\mathcal{S}_{\hat{g}}\ne \mathcal{S}_{\check{g}}, ~~ \forall \hat{g}\ne \check{g},
    \end{align}
where $\mathcal{S}_g$ stands for the commonly shared support set for the $g$-th group.

From Definition~1, it is worth noting that 1) $\w_k^s$ indicates which group the $k$-th MU belongs to; and 2)
the additional individual  sparse representation
vector $\w_k^v$  accounts for any outliers.
Clearly, the general sparsity model is more reasonable in practical implementations. It includes the commonly shared sparsity as a  special case with $\w_k^v=0$. Moreover, the outlier problem in the scenario of a single group has been addressed in \cite{rao2014distributed}, where the named joint sparsity model used  can also be included as a special case of ours with $G=1$.
Fig.~\ref{figMIMO} shows an example of the general sparsity model where the commonly shared support sets for two groups are $\mathcal{S}_1=\{1,2,3\}$ and $\mathcal{S}_2=\{7,8,10\}$, respectively. Specifically, if $\w_1=[\times,\times,\times,\times,0,0,\times,0,0,0]^T$, with \lq\lq$\times$" standing for a significant value, its corresponding general sparsity pattern is $\w_1^s=[\times,\times,\times,0,0,0,0,0,0,0]^T$ and $\w_1^v=[0,0,0,\times,0,0,\times,0,0,0]^T$.

%
%


The aim of this paper is to automatically partition the users into $G$ groups according to their hidden commonly shared
supports (determined by $\w_k^s$), and simultaneously obtain the channel estimation for each user.
This expected to obtain more accurate channel estimation performance because we exploit additional information about common sparsity among nearby users, as captured by (\ref{eq-parss22}).
The main challenge introduced by the general sparsity model is that it is difficult to directly extract the  commonly shared component $\w_k^s$ from $\w_k$ because $\w_k^s$ and $\w_k^v$ are mixed.
To the best of our knowledge, there lacks an efficient method for simultaneously handling $\w_k^s$ and $\w_k^v$.
In the presence of a uniform sparsity model (i.e., $\w_k^v=\mathbf{0}, \forall k$), the proposed method for wideband DOA estimation in \cite{wang2016novel}  is the only candidate that may be
tailored to solve the problem of joint channel estimation and user grouping. However, it does not apply to the more practical general sparsity model.
To overcome the challenge, in the next section, we propose a novel SBL-based framework which can fully exploit the common sparsity among nearby users and exclude the harmful effect from outliers simultaneously.



\noindent\textbf{Remark 2.} Although we set the number of groups $G$ to a fixed value in the general sparsity model, this fixed value is not required to be the real number of user groups $G^\star$.
When $G$ is chosen to be larger than $G^\star$, the algorithm will automatically cluster users into $G^\star$ groups and
assign zero users to the remaining $G-G^\star$ groups. Therefore, the number of groups can be \lq\lq optimized" by the proposed algorithm in an implicit way. When $G$ is smaller than the optimal $G^\star$, the outliers deviated from the uniform sparsity pattern can be mitigated by the general model. As a result, the channel estimation performance of the proposed algorithm is not sensitive to the choice of $G$ (which will be verified in the simulations).


\section{Joint Channel Estimation and User Grouping}

In this section, we propose an efficient SBL-based method for joint channel estimation and user grouping with the general sparsity model.
For ease of exposition, we proceed as follows. We begin by introducing the SBL formulation for group sparse signal recovery. Then, we  resort to the variational Bayesian inference (VBI) methodology \cite{tzikas2008variational} and adopt an alternating optimization algorithm to perform the Bayesian inference, so as to jointly cluster the users and estimate the channel. Note that the modeling error caused by direction mismatch will be addressed in the next section.


\subsection{Sparse Bayesian Learning Formulation}

In order to separate the commonly shared support and the individual support for the $k$-th MU, we use (\ref{eq-modpss}) to rewrite the received signal $\y_k$ as
\begin{align}\label{eq_re2}
\y_k=  \bm\Phi (\w_k^s + \w_k^v) + \n_k=  \bar{\bm\Phi}\bar\w_k+ \n_k,
\end{align}
where $\bar{\bm\Phi}=[\bm\Phi, \bm\Phi]$ and $\bar\w_k=[  (\w^s_k)^T, (\w^v_k)^T ]^T$.
Following the classical sparse Bayesian model \cite{tipping2001sparse}, we model $\w_k^s$ and $\w_k^v$ associated with user $k$ in group $g$ as
non-stationary Gaussian prior distributions:
\begin{align}
p(\w_k^s|\bm{\gamma}^*_g)=  \mathcal{CN}(\w_k^s | \bm{0}, \mathrm{diag}\left(\bm{\gamma}^*_g\right)^{-1} ),  \forall k\in \mathcal{G}_g \label{eq:mos}
\end{align}
and
\begin{align}\label{eqsspor}
p(\w_k^v| \bm{\gamma}^v_k)= \mathcal{CN} ( \w_k^v | \bm{0}, \rho\cdot \mathrm{diag}\left(\bm{\gamma}^v_k\right)^{-1}),
\end{align}
where  $\rho$ is a small positive constant (whose function will be explained later), $\bm{\gamma}^*_g =[\gamma^*_{g,1}, \gamma^*_{g,2},\ldots, \gamma^*_{g,\hat{L}}]^T$, $\bm{\gamma}_k^v =[\gamma^v_{k,1}, \gamma^v_{k,2},\ldots, \gamma^v_{k,\hat{L}}]^T$, and $\gamma^*_{g,l}$ and $\gamma^v_{k,l}$ stand for the precision of the $l$-th elements of $\w^s_k$ and $\w^v_k$, respectively.
Note that all users in group $g$ share a common precision vector $\bm\gamma^*_g$ for the common sparse vector $\w^s_k$, which captures the common sparsity shared by the users. On the other hand, different users in group $g$ have different precision vectors $\bm\gamma_k^s$ for the individual sparse vector $\w^v_k$, which captures the individual sparsity caused by the outliers deviated from the uniform sparsity pattern.
For a given sparse vector $\w_k$, there are multiple ways to partition $\w_k$ into a common sparse vector $\w_k^s$ and an individual sparse vector $\w_k^v$, where each partition corresponds to one user grouping result. Clearly, a user grouping result is only meaningful when the users in the same user group share a large common support; i.e., we favor a denser $\w_k^s$ over $\w_k^v$.
Hence, we introduce a small positive constant $\rho\in (0,1)$  in (\ref{eqsspor}) to provide a sparser prior  for $\w_k^v$ than for $\w_k^s$. Empirical evidence shows that the performance of our method is not sensitive to the choice of $\rho$, as long as $\rho$ is sufficiently small. In the simulations, we set $\rho=0.001$.


To force the clustering of $\w_k^s$s with $G$ groups,  we introduce $\z_k$ of size $G\times 1$ as the assignment vector for the $k$-th MU. Specifically, if the $k$-th MU belongs to the $g$-th group (i.e., $k\in \mathcal{G}_g$),
$\z_k$ is a zero vector, except for the $g$-th element being one.
Then, the distribution of $\w_k^s$ conditional on $\z_k$ and $\bm\gamma_{g}^* $s can be expressed as
\begin{align}\label{eqpgou}
p(\w_k^s | \z_k, \bm\Gamma^* )  = \prod_{g=1}^G \left\{ \mathcal{CN} ( \w_k^s | \bm{0}, \mathrm{diag}(\bm{\gamma}_{g}^*  )^{-1}  \right\}^{z_{k,g}},
\end{align}
where $\bm\Gamma^* =\{\bm\gamma^*_g\}_{g=1}^G$, and 
$z_{k,g}$ stands for  the $g$-th element of $\z_k$.

For tractable inference of $\bm\gamma^*_g$s and $\bm\gamma^v_k$s, the elements of $\bm\gamma^*_g$ and $\bm\gamma^v_k$ (denoted by $\gamma^*_{g,l}$ and $\gamma^v_{k,l}$, $l=1,2,\ldots,\hat L$)  are modeled as independent Gamma distributions, i.e.,
\begin{align}\label{eqsga}
p(\bm\gamma_g^*)=& \prod_{l=1}^{\hat L} \Gamma (\gamma^*_{g,l}| a, b)
\end{align}
and
\begin{align}\label{eqsgav}
p(\bm\gamma_k^v)=& \prod_{l=1}^{\hat L} \Gamma (\gamma^v_{k,l}| a, b),
\end{align}
where $a$ and $b$ are some small constants (e.g., $a=b=0.0001$). 
Gamma distribution is a conjugate prior of Gaussian distribution, and
the two-stage hierarchical prior provided by (\ref{eqpgou}) and (\ref{eqsga}) [or (\ref{eqsspor}) and (\ref{eqsgav})] for $\w_k^s$ (or $\w_k^v$) is recognized as encouraging sparsity, due to the heavy tails and sharp peak at zero \cite{tipping2001sparse,wipf2004sparse}.
In fact, it can be shown that finding a MAP estimate of $\w_k^v$ (or $\w_k^s$) with the two-stage hierarchical prior is equivalent to  finding the minimum $l_0$-norm solution using FOCUSS with $p\rightarrow0$ \cite{rao1999affine}, where $p$ corresponds to the parameter of $l_p$-norm.
It is worth noting that the precisions $\gamma_{g,l}^*$s in (\ref{eq:mos}) [or $\gamma_{k,l}^v$s in (\ref{eqsspor})] directly indicate the support of $\w^s_k$ (or $\w^v_k$). For example, if $\gamma_{g,l}^*$ is large, the $l$-th element of $\w^s_k$ tends to zero; otherwise, the value of the $l$-th element is significant.


Under the assumption of circular symmetric complex Gaussian noise, we have
\begin{align}\label{eq-yat}
p(\y_k | \w_k^s, \w_k^v, \alpha) =\mathcal{CN}(\y_k | \bm\Phi\w_k, \alpha^{-1}\I),
\end{align}
where $\alpha= \sigma^{-2}$ stands for the noise precision. Since $\alpha$ is usually unknown, we similarly model it as a gamma hyperprior $p(\alpha)=\Gamma(\alpha|a,b)$.

Let $\bm\Theta=\{ \alpha, \bar\W, \bm\Gamma^*, \bm\Gamma^v, \Z,\} $ be the set of hidden variables to be estimated, where
$\bar\W=\{\bar\w_k\}_{k=1}^K$,  $\bm\Gamma^v=\{\bm\gamma_k^v\}_{k=1}^K$, and $\Z=\{\z_k\}_{k=1}^K$.
The user groups and channel estimation can be jointly obtained if we can calculate  the maximum a $posteriori$ (MAP) optimal estimate of $p(\bm\Theta| \Y)$, where  $\Y=\{\y_k \}_{k=1}^K$.
Specifically, the user group is indicated by the MAP estimator of the group assignment vector $\z_k$, and the angular domain channel vector $\w_k$ can be calculated from the MAP estimator of $\bar\w_k$ according to (\ref{eq-modpss}).
Unfortunately, this MAP estimate is intractable. Therefore, in the next subsection, we will resort to the VBI methodology  and will adopt an alternating optimization algorithm to infer the hidden variables iteratively.


\subsection{Overview of the Proposed Method}

\begin{figure}
\center
\includegraphics[scale=0.3]{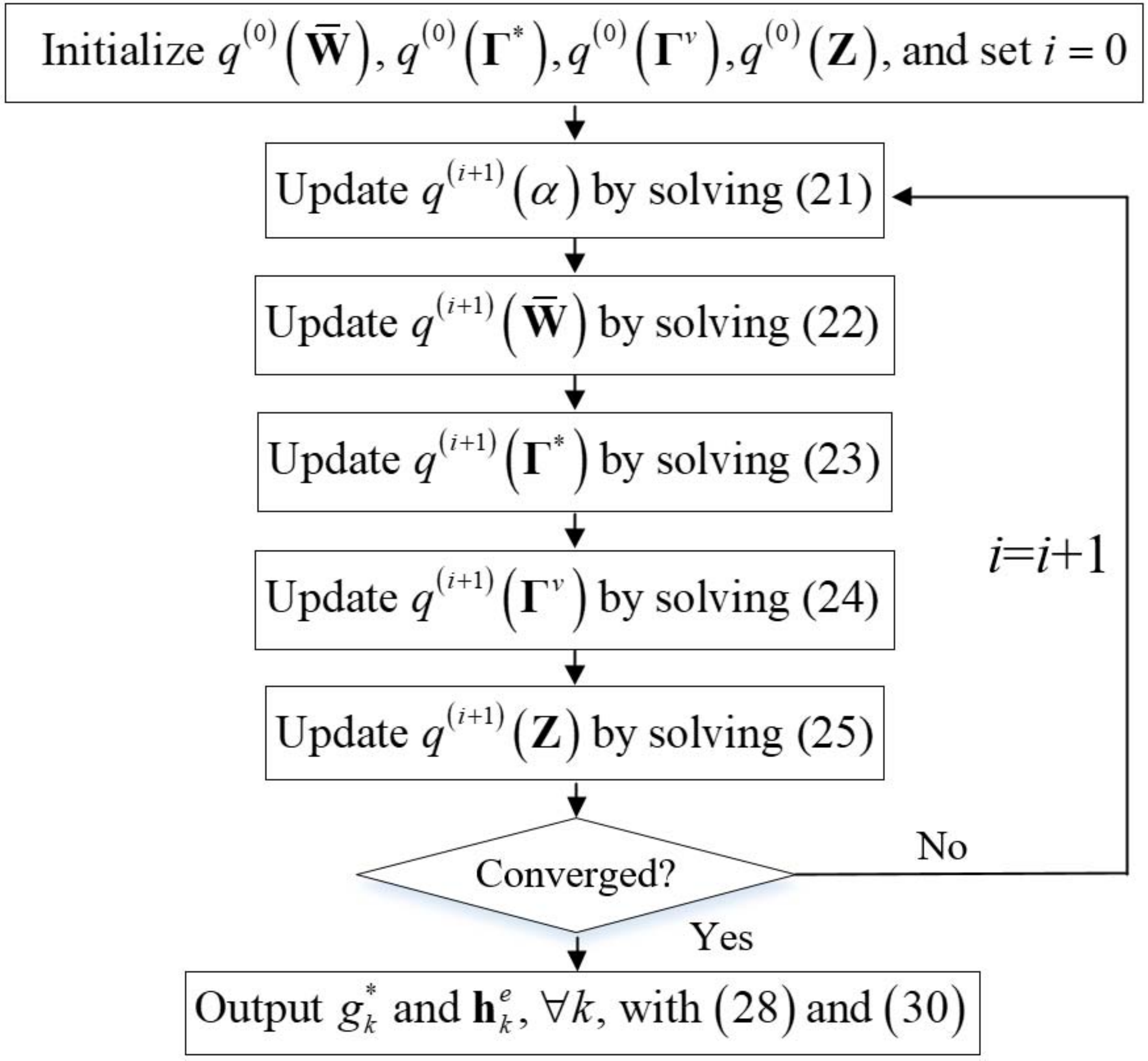}
\caption{The overall flow of the proposed algorithm.}\label{figFlow}
\end{figure}

The principle behind  VBI is to find an approximate posterior of $\bm\Theta$ (denoted by $q(\bm\Theta)$), instead of the exact posterior, to make the MAP estimate tractable, where $q(\bm\Theta)$ is assumed to be factorized approximately as
\begin{align}\label{eq-fra}
q(\bm\Theta )
=&  q(\alpha)\underbrace{ \prod_{k=1}^K q(\bar\w_k)}_{\triangleq q(\bar\W)} \underbrace{\prod_{g=1}^G q(\bm\gamma^*_g)}_{\triangleq q(\bm\Gamma^*)}\underbrace{\prod_{k=1}^K q(\bm\gamma^v_k)}_{\triangleq q(\bm\Gamma^v)}  \underbrace{\prod_{k=1}^K q(\z_k)}_{\triangleq q(\Z)},
\end{align}
and it should be chosen to minimize the Kullback-Leibler (KL) divergence with respect to (w.r.t.) the true posterior:
\begin{align}
D_{KL}(q(\bm\Theta )|| p(\bm\Theta| \Y)  )
=- \int  q(\bm\Theta )  \ln \frac{ p(\bm\Theta| \Y)}{q(\bm\Theta )} d\bm\Theta.
\end{align}
In other words, the corresponding optimization problem to find the \lq\lq best" approximate posterior under the factorized constraint in (\ref{eq-fra}) can be formulated as
\begin{align}\label{eq-all1}
q^\star(\bm\Theta )= \arg\max_{q(\bm\Theta )}  \underbrace{\int  q(\bm\Theta )  \ln \frac{ p( \Y, \bm\Theta)}{q(\bm\Theta )} d\bm\Theta}_{\triangleq \mathcal{U}( q_{1}, q_{2},q_{3},q_{4},q_{5}  )},
\end{align}
where $q_{i}$ denotes $q(\Theta_i)$ for simplicity, and $\Theta_i$ stands for the $i$-th element in $\bm\Theta$.
Since the above objective is a high-dimensional non-convex function, it is difficult to find the optimal solution. Here, we adopt an alternating optimization algorithm to find a stationary solution instead. Specifically, we update $q_i$s as
\begin{align}
q_1^{(i+1)}&= \arg \max_{q_1}   \mathcal{U}(  q_{1}, q_{2}^{(i)},q_{3}^{(i)},q_{4}^{(i)},q_{5}^{(i)}   ),\label{eqM1}\\
q_2^{(i+1)}&= \arg \max_{q_2}   \mathcal{U}(  q_{1}^{(i+1)}, q_{2},q_{3}^{(i)},q_{4}^{(i)},q_{5}^{(i)}   ),\label{eqM2}\\
q_3^{(i+1)}&= \arg \max_{q_3}   \mathcal{U}(  q_{1}^{(i+1)}, q_{2}^{(i+1)},q_{3},q_{4}^{(i)},q_{5}^{(i)}   ),\label{eqM3}\\
q_4^{(i+1)}&= \arg \max_{q_4}   \mathcal{U}(  q_{1}^{(i+1)}, q_{2}^{(i+1)},q_{3}^{(i+1)},q_{4},q_{5}^{(i)}   ),\label{eqM4}\\
q_5^{(i+1)}&= \arg \max_{q_5}   \mathcal{U}(  q_{1}^{(i+1)}, q_{2}^{(i+1)},q_{3}^{(i+1)},q_{4}^{(i+1)},q_{5}   ),\label{eqM5}
\end{align}
where $(\cdot)^{(i)}$ stands for the $i$-th iteration.
Once the algorithm converges, we can obtain the approximate posteriors: $q(\alpha)$, $q(\bar\w_k)$s, $q(\bm\gamma^*_g)$s, $q(\bm\gamma^v_k)$s and  $q(\z_k)$s.
Let
\begin{align}
\hat{\bm\phi}_{k}=&  \left< \z_{k} \right>_{q(\z_k)}
\end{align}
and
\begin{align}
\bar{\bm\mu}_k =&  \left< \bar{\w}_k \right>_{q(\bar{\w}_k)},
\end{align}
where $\left< \cdot \right>_{p(x)}$ stands for the expectation operator w.r.t. $p(x)$.
Then, we are able to cluster the users into $G$ groups; e.g., user $k$ belongs to group $g_k^\star$ if
\begin{align}
g_k^\star=\arg\max_g  \hat\phi_{k,g},
\end{align}
where $\hat\phi_{k,g}$ stands for the $g$-th element of $\hat{\bm\phi}_{k}$. Recall that $\w_k=\w_k^s +\w_k^v$ and $\bar\w_k=[  (\w^s_k)^T, (\w^v_k)^T ]^T$. Therefore, we have
\begin{align}
\bm\mu_k \triangleq&  \left<   \w_k  \right>_{q(\bar\w_k)}= \bar{\bm\mu}_{k,1} + \bar{\bm\mu}_{k,2},
\end{align}
where $ \bar{\bm\mu}_{k,1}$ and $\bar{\bm\mu}_{k,2}$ stand for the first and last $\hat L$ elements of $\bar{\bm\mu}_k$, respectively.
Letting $\Omega_k=\mathrm{supp}(\bm\mu_k)$, the estimated downlink channels $\h_k^e$s can be calculated by
\begin{align}
\h_k^e= \A_{\Omega_k} \left(\bm\Phi_{\Omega_k}\right)^{\dag} \y_k.
\end{align}
The overall flow of the proposed algorithm is given in Fig.~\ref{figFlow}.
In the following subsections, we will illustrate how to solve the optimization problems (\ref{eqM1})--(\ref{eqM5}) in detail (Section~III-C)
and then give a convergence analysis of the alternating optimization algorithm (Section~III-D).

\subsection{Detailed Implementations}

%
%
%

In this subsection, we focus on handling the optimization problems (\ref{eqM1})--(\ref{eqM5}), whose solutions will be presented in Lemmas 3--7 as follows.
It is worth noting that some initializations are required to trigger the iterations, which will be addressed later.

\subsubsection{\underline{Update for $q_1$}} We update $q_1$ [or $q(\alpha)$] by solving the optimization problem (\ref{eqM1}), whose solution follows a gamma distribution.

\noindent\textbf{Lemma~3.}  The optimization problem (\ref{eqM1}) has a unique solution:
\begin{align}\label{eqle1}
q^{(i+1)}(\alpha)=& \Gamma(\alpha |  a_\alpha^{(i)}, b_\alpha^{(i)}  ),
\end{align}
where $ a_\alpha^{(i)}= (a + KT ) $ and $b_\alpha^{(i)} = b+ \sum_{k=1}^K (\| \y_k - \bm\Phi \bm\mu_k^{(i)}    \|_2^2     + \tr(\bm\Phi \bm\Sigma_k^{(i)}\bm\Phi^H))$, with $\bm\mu_k^{(i)}$ and $\bm\Sigma_k^{(i)} $  being the mean and variance of $\w_k$ at the $i$-th iteration [whose closed-from expressions will be given later, in (\ref{eqclmu}) and (\ref{eqclSigma})].
\begin{proof}
See Appendix\ref{lemmaa}.
\end{proof}


Note that the mean of $\alpha$, w.r.t. the gamma distribution defined in (\ref{eqle1}),  can be calculated as a byproduct:
\begin{align}
\hat{\alpha}^{(i+1)}\triangleq \left< \alpha \right>_{q^{(i+1)}(\alpha)}   =\frac{ a_\alpha^{(i)}}{ b_\alpha^{(i)} },
\end{align}
which will be used in the next lemma.

\subsubsection{\underline{Update for $q_2$}} We update $q_2$ [or $q(\bar\W)$] by solving the optimization problem (\ref{eqM2}), whose solution follows a Gaussian distribution.

\noindent\textbf{Lemma~4.}  The optimization problem (\ref{eqM2}) has a unique solution:
\begin{align}\label{eqle1-2}
q^{(i+1)}(\bar\W) = \prod_{k=1}^Kq^{(i+1)}(\bar\w_k) =\prod_{k=1}^K\mathcal{CN}(\bar\w_k | \bar{\bm\mu}_k^{(i+1)}, \bar{\bm\Sigma}_k^{(i+1)} ),
\end{align}
where $\bar{\bm\mu}_k^{(i+1)}= \hat{\alpha}^{(i+1)} \bar{\bm\Sigma}_k^{(i+1)} \bar{\bm\Phi}^H  \y_k  $ and $\bar{\bm\Sigma}_k^{(i+1)}  =  \left(\hat{\alpha}^{(i+1)} \bar{\bm\Phi}^H \bar{\bm\Phi}  + \diag([(\hat{\bm\gamma}^{s}_k)^{(i)};
\rho^{-1}(\hat{\bm\gamma}^{v}_k)^{(i)}])\right)^{-1}$ with $(\hat{\bm\gamma}^{s}_k)^{(i)}$ and $(\hat{\bm\gamma}^{v}_k)^{(i)}$ being the means of $\bm\gamma^{s}_k$ and $\bm\gamma^{v}_k$ at the $i$-th iteration (whose closed-from expressions will be given later, in (\ref{equpss2}) and (\ref{eqqvg2}), respectively).
\begin{proof}
See Appendix\ref{lemmab}.
\end{proof}

Since $\w_k= \w_k^s + \w_k^v$, we have
\begin{align}
\bm\mu_k^{(i+1)} \triangleq&  \left<   \w_k  \right>_{q^{(i+1)}(\bar\w_k)}= \bar{\bm\mu}_{k,1}^{(i+1)} + \bar{\bm\mu}_{k,2}^{(i+1)}\label{eqclmu}
\end{align}
and
\begin{align}
\bm\Sigma_k^{(i+1)} \triangleq&  \left<  ({\w}_k - \bm\mu_k^{(i+1)} )({\w}_k - \bm\mu_k^{(i+1)} )^H\right>_{q^{(i+1)}(\bar\w_k)}\notag\\
=& \bar{\bm\Sigma}_{k,1}^{(i+1)} + \bar{\bm\Sigma}_{k,2}^{(i+1)} + \bar{\bm\Sigma}_{k,3}^{(i+1)}+ \bar{\bm\Sigma}_{k,4}^{(i+1)},\label{eqclSigma}
\end{align}
where $\bar{\bm\mu}_{k,1}^{(i+1)}= \bar{\bm\mu}_{k}^{(i+1)}(1:\hat L)$, $\bar{\bm\mu}_{k,2}^{(i+1)}= \bar{\bm\mu}_{k}^{(i+1)}(\hat L+1: 2\hat L)$, $\bar{\bm\Sigma}_{k,1}^{(i+1)}=\bar{\bm\Sigma}_{k}^{(i+1)}(1:\hat L, 1:\hat L)$,
$\bar{\bm\Sigma}_{k,2}^{(i+1)}=\bar{\bm\Sigma}_{k}^{(i+1)}(\hat L+1: 2\hat L, \hat L+1: 2\hat L)$,
$\bar{\bm\Sigma}_{k,3}^{(i+1)}=\bar{\bm\Sigma}_{k}^{(i+1)}(1:\hat L, \hat L+1: 2\hat L)$,
and $\bar{\bm\Sigma}_{k,4}^{(i+1)}=\bar{\bm\Sigma}_{k}^{(i+1)}(\hat L+1: 2\hat L, 1:\hat L)$. Note that these byproducts will be required for updating both  $q_3$ and $q_4$.

\subsubsection{\underline{Update for $q_3$}} We update $q_3$ [or $q(\bm\Gamma^*)$] by solving the optimization problem (\ref{eqM3}), whose solution follows a gamma distribution.

\noindent\textbf{Lemma~5.}  The optimization problem (\ref{eqM3}) has a unique solution:
\begin{align}\label{eqle1-3}
q^{(i+1)}(\bm\Gamma^*) =   & \prod_{g=1}^G\prod_{l=1}^{\hat L} q^{(i+1)}(\gamma^*_{g,l})\notag\\
=& \prod_{g=1}^G\prod_{l=1}^{\hat L}   \Gamma\left(\gamma^*_{g,l} | (a^*_{g,l})^{(i+1)}, (b^*_{g,l})^{(i+1)}\right),
\end{align}
where $(a^*_{g,l})^{(i+1)}= a +  \sum_{k=1}^K\hat\phi_{k,g}^{(i)}$,  $ (b^*_{g,l})^{(i+1)}= b +  \sum_{k=1}^K  \hat\phi_{k,g}^{(i)} ( |\bar{\mu}_{k,1,l}^{(i+1)}|^2 +\bar{\Sigma}_{k,1,l}^{(i+1)}  )$,
$\hat\phi_{k,g}^{(i)}= q^{(i)} (z_{k,g}=1)$ [whose closed-from expression will be given later, in (\ref{equpz1})],
$\bar{\mu}_{k,1,l}^{(i+1)}$ stands for the $l$-th element of $\bar{\bm\mu}_{k,1}^{(i+1)}$, and  $\bar{\Sigma}_{k,1,l}^{(i+1)}$ stands for the $l$-th diagonal element of $\bar{\bm\Sigma}_{k,1}^{(i+1)}$.
\begin{proof}
See Appendix\ref{lemmac}.
\end{proof}

Then, the mean of $\gamma^*_{g,l}$ at the $(i+1)$-th iteration is
\begin{align}
(\hat{\gamma}^*_{g,l})^{(i+1)}\triangleq&\left<\gamma^*_{g,l}\right>_{q^{(i+1)}(\gamma^*_{g,l})}=  \frac{   (a^*_{g,l})^{(i+1)} }{\left((b^*_{g,l})^{(i+1)}\right)},
\end{align}
and the logarithmic expectation is
\begin{align}
(\widehat{\ln{\gamma}^*_{g,l}})^{(i+1)} \triangleq&\left<\ln \gamma^*_{g,l}\right>_{q^{(i+1)}(\gamma^*_{g,l})}\notag\\
=&  \Psi \left(  (a^*_{g,l})^{(i+1)}  \right)  - \ln \left((b^*_{g,l})^{(i+1)}\right),\label{equpss3}
\end{align}
where $\Psi(\cdot)$ stands for the digamma function.
We define ${\gamma}^{s}_{k,l}= \sum_{g=1}^G  \left( z_{k,g}\gamma^*_{g,l}  \right)  $, and then the mean of ${\gamma}^{s}_{k,l}$ at the $(i+1)$-th iteration is
\begin{align}
(\hat{\gamma}^s_{k,l})^{(i+1)}=&\left<\gamma^s_{k,l}\right>_{q^{(i+1)}(\bm\Gamma^*)q^{(i)}(\z_k)}=  \sum_{g=1}^G   \hat\phi^{(i)}_{k,g}(\hat{\gamma}^{*}_{g,l})^{(i+1)}.\label{equpss2}
\end{align}
Note that (\ref{equpss2})
will be required for updating $q(\Z)$.

\subsubsection{\underline{Update for $q_4$}} We update $q_4$ [or $q(\bm\Gamma^v)$] by solving the optimization problem (\ref{eqM4}), whose solution also follows a gamma distribution.

\noindent\textbf{Lemma~6.}  The optimization problem (\ref{eqM4}) has a unique solution:
\begin{align}
q^{(i+1)}(\bm\Gamma^v) =   & \prod_{k=1}^K\prod_{l=1}^{\hat L} q^{(i+1)}(\gamma^v_{k,l})\notag\\
=& \prod_{k=1}^K\prod_{l=1}^{\hat L}   \Gamma\left(\gamma^v_{k,l} | (a^v_{k,l})^{(i+1)}, (b^v_{k,l})^{(i+1)}\right),\label{eqqvg}
\end{align}
where $(a^v_{k,l})^{(i+1)}=a + K $  and $(b^v_{k,l})^{(i+1)}=  b +  \rho^{-1} ( |\bar{\mu}_{k,2,l}^{(i+1)}|^2 +\bar{\Sigma}_{k,2,l}^{(i+1)}  )$.
\begin{proof}
The proof is similar to Lemma~3. So it is omitted for brevity.
\end{proof}

With (\ref{eqqvg}), the mean of $\gamma^v_{k,l}$ at the $(i+1)$-th iteration is
\begin{align}
(\hat{\gamma}^v_{k,l})^{(i+1)}\triangleq&\left<\gamma^v_{k,l}\right>_{q^{(i+1)}(\gamma^v_{k,l})}=  \frac{   (a^v_{k,l})^{(i+1)} }{\left((b^v_{k,l})^{(i+1)}\right)},\label{eqqvg2}
\end{align}
which was required for updating $q_2$.

\subsubsection{\underline{Update for $q_5$}} We update $q_5$ [or $q(\Z)$] by solving the optimization problem (\ref{eqM5}), whose solution is characterized by the following lemma.

\noindent\textbf{Lemma~7.}  The optimization problem (\ref{eqM5}) has a unique solution:
\begin{align}\label{eqle1-5}
q^{(i+1)}(\Z) = \prod_{k=1}^K q^{(i+1)}(\z_k)= \prod_{k=1}^K    \prod_{g=1}^G \left(\hat\phi_{k,g} ^{(i+1)}  \right)^{z_{k,g}},
\end{align}
where
\begin{align}
\hat\phi_{k,g}^{(i+1)}= q^{(i+1)} (z_{k,g}=1)  =  \frac{\exp(\varsigma_{k,g}^{(i+1)})}{\sum_{g=1}^G \exp( \varsigma_{k,g}^{(i+1)}  )}\label{equpz1}
\end{align}
and
\begin{small}
\begin{align}
\varsigma_{k,g}^{(i+1)}={\sum_{l=1}^{\hat L} (\widehat{\ln \gamma^*_{g,l}})^{(i+1)} -
\sum_{l=1}^{\hat L} (\gamma^*_{g,l})^{(i+1)}\left( |\bar{\mu}_{k,1,l}^{(i+1)}|^2 +\bar{\Sigma}_{k,1,l}^{(i+1)}  \right)}.
\end{align}
\end{small}
\begin{proof}
See Appendix\ref{lemmad}.
\end{proof}



The proposed algorithm proceeds by repeated application of (\ref{eqle1}), (\ref{eqle1-2}), (\ref{eqle1-3}), (\ref{eqqvg}) and (\ref{eqle1-5}), and its convergence will be addressed in the next subsection.
The main computational burden of the proposed
method is given as follows.
\begin{itemize}
  \item The most cost for updating $q(\alpha)$ is to calculate $b_\alpha$, whose computational complexity is $\mathcal{O}(T \hat{L}^2 K)$  per iteration.
  \item
  Calculating  $\bar{\bm\Sigma}_k$s and $\bar{\bm\mu}_k$s in each iteration for updating $q(\bar{\W})$ is $\mathcal{O}(T\hat{L}^2K)$ and $\mathcal{O}(\hat{L}^2K)$, respectively.
  \item The complexities in updating $q(\bm\Gamma^\star)$ and $q(\bm\Gamma^v)$ in each iteration are $\mathcal{O}( G\hat{L} K  )$ and $\mathcal{O}( \hat{L} K  )$, respectively.
   \item The complexity in updating $q(\Z)$ is $\mathcal{O}(G\hat{L})K$  per iteration.
\end{itemize}
This suggests the total computational requirement of the proposed method is  $\mathcal{O}(T \hat{L}^2 K)$ per iteration.

Following are some practical implementation tips for the proposed method.
In order to trigger the alternating optimization algorithm, we need some initializations for $q^{(0)}(\bar\W)$, $q^{(0)}(\bm\Gamma^*)$, $q^{(0)}(\bm\Gamma^v)$ and $q^{(0)}(\Z)$.
According to the main results in Lemmas 4--7, we can simply set the initializations as follows:
\begin{itemize}
  \item  initialize a Gaussian distribution function $q^{(0)}(\bar\W)$,
with parameters $\bar{\bm\mu}_k^{(i+1)}= \bar{\bm\Sigma}_k^{(0)} \bar{\bm\Phi}^H  \y_k  $ and $\bar{\bm\Sigma}_k^{(0)}  =  (\bar{\bm\Phi}^H \bar{\bm\Phi}  + \diag([\bm{1}_{\hat L\times1};
\rho^{-1} \bm{1}_{\hat L\times1}]  ))^{-1}$;
  \item initialize a gamma distribution function $q^{(0)}(\bm\Gamma^*)$, with parameters $(a^*_{g,l})^{(0)}=(b^*_{g,l})^{(0)}=1, \forall g,l$;
  \item initialize a gamma distribution function $q^{(0)}(\bm\Gamma^v)$, with parameters $(a^v_{k,l})^{(0)}=(b^v_{k,l})^{(0)}= 1, \forall k,l$;
  \item initialize  $q^{(0)}(\Z)$, with $\varsigma_{k,g}^{(0)}$s being uniformly chosen from $[0,1]$.
\end{itemize}
Empirical evidence shows that the proposed method remains very robust to these initializations. Moreover, we set
$a=b=0.0001$ in the simulations.

%

\subsection{Convergence Analysis and Discussion}

The non-decreasing property of the sequence  $\mathcal{U}(  q_{1}^{(i)}, q_{2}^{(i)},q_{3}^{(i)},q_{4}^{(i)},q_{5}^{(i)}   )$, $i=1,2,3,\ldots$, is well guaranteed by the update rules (\ref{eqM1})--(\ref{eqM5}).

\noindent\textbf{Lemma~8.} The update rules (\ref{eqM1})--(\ref{eqM5}) give a non-decreasing sequence  $\mathcal{U}(  q_{1}^{(i)}, q_{2}^{(i)},q_{3}^{(i)},q_{4}^{(i)},q_{5}^{(i)}   )$, $i=1,2,3,\ldots$.
\begin{proof}
See Appendix\ref{lemmae}.
\end{proof}

Together with the fact that the objective function $\mathcal{U}(  q_{1}, q_{2},q_{3},q_{4},q_{5}  )$  has an upper bound of $1$,\footnote{
This is because of $\int  q(\bm\Theta )  \ln \frac{ p( \Y, \bm\Theta)}{q(\bm\Theta )} d\bm\Theta\le \ln\int  q(\bm\Theta )  \frac{ p( \Y, \bm\Theta)}{q(\bm\Theta )} d\bm\Theta=\ln p(\Y)$,
where the first inequality follows Jensen's inequality.} the sequence  $\mathcal{U}(  q_{1}^{(i)}, q_{2}^{(i)},q_{3}^{(i)},q_{4}^{(i)},q_{5}^{(i)}   )$, $i=1,2,3,\ldots$, converges to a limit.
The alternating algorithm does not converge to a stationary solution in general. However, the specific conditions satisfied by our problem make it possible to prove the convergence of the alternating algorithm to a stationary point.
The alternating optimization algorithm  can be viewed as a special case of the block MM algorithm.  Hence, we have the following lemma:

\noindent\textbf{Lemma~9.} The iterates generated by the alternating optimization algorithm converge to a stationary solution of the optimization problem (\ref{eq-all1}).
\begin{proof}
See Appendix\ref{lemmaStationary}.
\end{proof}

Finally, we discuss  the relationship  between our method and the method in \cite{wang2016novel}:
\begin{itemize}
  \item Recall that the general sparsity model used in our method includes the commonly shared sparsity model used in \cite{wang2016novel}
  as a special case of $\w_k^v=\mathbf{0},\forall k$. Thereore, our method designed for the general sparsity model is more  general than the method in \cite{wang2016novel}. It can also handle the commonly shared sparsity model, by simply ignoring the updates for $\w_k^v$s and $\bm\gamma_k^v$s.

  \item Our method performs Bayesian inference for the hidden variables from a new perspective of alternating optimization.
   Compared with the traditional Bayesian inference used in \cite{wang2016novel}, our method has the following advantages:
   1) its convergence is more easily proved (see Lemma~8); 2) it reveals that the convergence solution is also a stationary solution (see Lemma~9), which is a stronger convergence result  since the traditional method only establishes the convergence of objective values to a certain point, without proving the converged solution is a stationary solution; and 3) it provides a flexible framework to handle the problem of direction mismatch (see Section IV).

  \item In our method,  each $\z_k$ is treated as a simple assignment vector without a prior distribution, and the number of groups $G$ is assumed to be known, while in \cite{wang2016novel}, each $\z_k$ is treated as a random vector 
      that is generated from a Dirichlet process prior, and the number of groups $G$ is automatically determined. It is worth noting that extending our method with the Dirichlet process (DP) prior and the automatically determined $G$ is straightforward. Even without such extending, empirical results (also refer to the simulations) show that our method is still  applicable to an unknown $G$. This is because the adopted general model can capture a much more general group sparse structure and can provide a robust result for an inexact choice of $G$ (Remark~2). The simulation results also show that there is no performance loss by removing the DP prior.



\end{itemize}

Another motivation for choosing a fixed $G$ comes from that fact that the user grouping result 
with a fixed $G$ can  be applied to some practical applications in massive MIMO systems. 
For example, we may combine the proposed method with Joint Spatial Division and Multiplexing (JSDM) \cite{adhikary2013joint,nam2014joint}, where a fixed $G$ is required.
It is worth noting that we do not try to
provide an improved JSDM framework, but only replace the user grouping algorithm used in JSDM with ours. This application is just a byproduct of our method.
Compared with the traditional user grouping method,  our method can bring some significant advantages: 1) it does not require prior knowledge about the channel covariance, where the acquisition of channel covariance may pose great challenges because it requires collecting  a large number of channel samples in practical implementations; and 2) it can give a better user grouping  result in the sense of Bayesian optimality, so as to alleviate the interference across different groups and enhance the sum-rate performance of JSDM systems.




\section{Handling Direction Mismatch with Arbitrary 2D-array Geometry}

In the section, we extend the proposed method to handling direction mismatch with an arbitrary 2D-array geometry. Note that the steering vector $\a(\theta,\phi)$ for an arbitrary 2D-array geometry contains both azimuth angle $\theta$ and elevation angle $\phi$ \cite{dai2018fdd,shen2018channel}:
\begin{align}
&\a(\theta,\phi)
=[1,  e^{-j2\pi  \frac{d_2}{\lambda}\cos(\phi)\sin(\theta- \psi_2) }, \notag\\
&~~~~~~~~~~~~~~~~~~~~~~~~\ldots, e^{-j2\pi  \frac{d_N }{\lambda}\cos(\phi)\sin(\theta- \psi_N)} ]^T,
\label{eq2Dsteer}
\end{align}
where $(d_n, \psi_n)$ is the coordinates of the $n$-th sensor.
Following the convention in Section III, we adopt a fixed sampling grid $\hat{\bm\vartheta}=\{\hat{\vartheta}_{l}\}_{l=1}^{\hat{L}}$ to uniformly cover the azimuth domain $[-\pi, \pi]$.
Recall that the direction mismatch between the true AoD and the grid point is unavoidable because signals usually come from random directions in practice.
Here, we adopt the off-grid model proposed in \cite{dai2018fdd} to handle the direction mismatch.
Let $\theta^k_l$ and $\phi^k_l$ denote the $l$-th true azimuth and elevation AoDs of the $k$-th MU, repectively.
If $\theta^k_l \notin \{\hat\vartheta_i\}_{i=1}^{\hat{L}}$ and $\hat\vartheta_{n_l}, n_l\in\{1,2,\ldots, \hat{L}\}$, is the nearest grid point to $\theta^k_l$,  we write $\theta^k_l$ as
\begin{align}
\theta^k_l= \hat\vartheta_{n_l} + \beta_{k,n_l},\label{eq-offg}
\end{align}
where $\beta_{k,n_l}$ corresponds to the azimuth direction  mismatch (or off-grid gap). With (\ref{eq-offg}), the received signal $\y_k$ can be rewritten by
\begin{align}\label{Hmodeln123}
\y_k= \bm\Phi(\bm\beta_k,\bm\varphi_k)(\w_k^s+ \w_k^v) + \n_k,
\end{align}
where $\bm\Phi(\bm\beta_k,{\bm\varphi}_k) = \X \A(\bm\beta_k,{\bm\varphi}_k) $,
$\bm\beta_k=[\beta_{k,1}, \beta_{k,2},\ldots, \beta_{k,\hat L}]^T$,
${\bm\varphi}_k=[{\varphi}_{k,1}, {\varphi}_{k,2},\ldots, {\varphi}_{k,\hat L}]^T$,
$\A(\bm\beta_k,{\bm\varphi}_k)=  [\a(\hat{\vartheta}_{1} +  \beta_{k,1},\varphi_{k,1}) ,\a(\hat{\vartheta}_{2} +  \beta_{k,2},\varphi_{k,2}),\ldots, \a(\hat{\vartheta}_{\hat L} +  \beta_{k, \hat L},\varphi_{k, \hat L})  ]$,
$\beta_{k,n_l}=\begin{cases}\theta^k_l - \hat\vartheta_{n_l},  &l=1,2,\ldots, L   \\
 0,   & \mathrm{otherwise}\end{cases}$, and
$\varphi_{k,n_l}=\begin{cases}\phi^k_l,  &l=1,2,\ldots, L   \\
 0,   & \mathrm{otherwise}\end{cases}$.
Note that $\varphi_{k,n_l}$ corresponds to  elevation direction mismatch.
Due to introducing the term of the off-grid gap, the direction mismatch can be significantly alleviated. Another advantage is that
the commonly shared support among nearby users does not need to coincide strictly with each other.
For example, let $\hat L=180$ and the  azimuth AoDs of two nearby MUs be $\{8.1^\circ,  10.2^\circ, 11.9^\circ , 15.3^\circ \}$ and $\{7.4^\circ,  10.3^\circ, 12.3^\circ , 15.1^\circ \}$, respectively. In this case, the nearest grid points for the first supports of the two MUs are different. However, an appropriate choice of off-grid gap can fix the commonly shared support mismatch, e.g.,
$8.1^\circ= 8^\circ + 0.1^\circ$ and  $7.4^\circ= 8^\circ - 0.6^\circ$.

In the sparse Bayesian  learning formulation for the off-grid model (\ref{Hmodeln123}), almost all the results in Section III-B remain unchanged, except that (\ref{eq-yat}) is replaced by
\begin{align}\label{eq-yat22}
p(\y_k | \w_k^s, \w_k^v, \alpha,\bm\beta_k,\bm\varphi_k) =\mathcal{CN}(\y_k | \bm\Phi(\bm\beta_k, \bm\varphi_k)\w_k, \alpha^{-1}\I)
\end{align}
and the optimization problem  (\ref{eq-all1}) is modified by
\begin{align}\label{eq-all2}
\left\{q^\star(\bm\Theta ), \B^\star  \right\}= \arg\max_{q(\bm\Theta ), \B}   \mathcal{U}( \bm\Theta, \B),
\end{align}
where $\B=\{\bm\beta_k, \bm\varphi_k\}_{k=1}^K$  is treated as a unknown parameter, rather than a random variable.
Similarly, in the $(i+1)$-th iteration,  we update  $q_i$s and $\B$ as
\begin{align}
q_1^{(i+1)}&= \arg \max_{q_1}   \mathcal{U}(  q_{1}, q_{2}^{(i)},q_{3}^{(i)},q_{4}^{(i)},q_{5}^{(i)},\B^{(i)},  ),\label{eq2M1}\\
q_2^{(i+1)}&= \arg \max_{q_2}   \mathcal{U}(  q_{1}^{(i+1)}, q_{2},q_{3}^{(i)},q_{4}^{(i)},q_{5}^{(i)},\B^{(i)},    ),\label{eq2M2}\\
q_3^{(i+1)}&= \arg \max_{q_3}   \mathcal{U}(  q_{1}^{(i+1)}, q_{2}^{(i+1)},q_{3},q_{4}^{(i)},q_{5}^{(i)},\B^{(i)},     ),\label{eq2M3}\\
q_4^{(i+1)}&= \arg \max_{q_4}   \mathcal{U}(  q_{1}^{(i+1)}, q_{2}^{(i+1)},q_{3}^{(i+1)},q_{4},q_{5}^{(i)},\B^{(i)},   ),\label{eq2M4}\\
q_5^{(i+1)}&= \arg \max_{q_5}   \mathcal{U}(  q_{1}^{(i+1)}, q_{2}^{(i+1)},q_{3}^{(i+1)},q_{4}^{(i+1)},q_{5},\B^{(i)}, ),\label{eq2M5}\\
\B^{(i+1)}&= \arg \max_{\B}   \mathcal{U}(  q_{1}^{(i+1)}, q_{2}^{(i+1)},q_{3}^{(i+1)},q_{4}^{(i+1)},q^{(i+1)}_{5},\B, ).\label{eq2M6}
\end{align}
Applying the results in Section III-C, we can obtain the solutions to (\ref{eq2M1})--(\ref{eq2M5}) directly, where the only difference is in replacing $\bm\Phi$ with $\bm\Phi(\bm\beta_k, \bm\varphi_k)$.

What remains is to obtain the update for $\B$. However, the last maximization problem (\ref{eq2M6}) is non-convex and it is difficult to find its optimal solution. Alternatively, we apply gradient update on the objective function of (\ref{eq2M6}) and  obtain a simple one-step update for each $\bm\beta_k$ and $\bm\varphi_k$ as in \cite{dai2018fdd}. As shown in  Appendix\ref{lemmaoff}, the derivative of the objective function, w.r.t. $\bm\beta_k$, can be calculated as
\begin{align}
\bm{\zeta}_k^{(i+1)}=[\zeta^{(i+1)}(\beta_{k,1}),\zeta^{(i+1)}(\beta_{k,2}),\ldots, \zeta^{(i+1)}(\beta_{k,\hat L})]^T, \label{eqderbeta}
\end{align}
with
\begin{align}
&\zeta^{(i+1)}(\beta_{k,l})\notag\\
=& 2 \mathrm{Re}\left(  (\a' (\hat\vartheta_{l}+ \beta_{k,l}, \varphi_{k,l}^{(i)}) )^H \X^H\X \a (\hat\vartheta_{l}+ \beta_{k,l},\varphi_{k,l}^{(i)}) \right) \cdot c_{k1}^{(i+1)}\notag\\
&+ 2 \mathrm{Re}\left(  (\a' (\hat\vartheta_{l}+ \beta_{k,l},\varphi_{k,l}^{(i)}))^H \X^H \c_{k2}^{(i+1)} \right), \label{eqderbeta1}
\end{align}
where $c_{k1}^{(i+1)}=-\hat{\alpha}^{(i+1)}(\chi_{k,ll}^{(i+1)}+ |\mu_{k,l}^{(i+1)}|^2) $, $\c_{k2}^{(i+1)}=\hat\alpha^{(i+1)} ((\mu_{k,l}^{(i+1)})^* \y_{k-l}^{(i+1)}   -\X \sum_{j\ne l} \chi_{k,jl}^{(i+1)} \a (\hat\vartheta_{j}+ \beta_{k,j}^{(i)},\varphi_{k,j}^{(i)}) )$,
$\y_{k-l}^{(i+1)}= \y_k -   \X \cdot\sum_{j\ne l } (\mu_{k,j}^{(i+1)} \cdot\a (\hat\vartheta_{j}+ \beta_{k,j}^{(i)},\varphi_{k,j}^{(i)})) $, $\a' (\hat\vartheta_{l}+\beta_{k,l},\varphi_{k,l})= d \a(\hat\vartheta_{l}+\beta_{k,l},\varphi_{k,l}) /{d \beta_{k,l}} $, and  $\mu_{k,l}^{(i+1)}$ and $\chi_{k,jl}^{(i+1)}$ denote the $l$-th element and the $(j,l)$-th element of $\bm\mu_{k}^{(i+1)}$ and $\bm\Sigma_{k}^{(i+1)}$, respectively.
With (\ref{eqderbeta}), we are able to update  the value of $\bm\beta_k$  in  the derivative direction, i.e.,
\begin{align}\label{equpv3}
\bm\beta_k^{(i+1)}= \bm\beta_k^{(i)}  + \Delta_{k} \cdot \bm{\zeta}_k^{(i+1)},
\end{align}
where $\Delta_{k}$ is the stepsize that can be optimized by  backtracking line search \cite{wright1999numerical}. As mentioned in Section III-D of \cite{dai2018fdd}, choosing the right stepsize can be time-consuming. To reduce the computational complexity, we use a fixed stepsize to update $\bm\beta_k$:
\begin{align}\label{equpv3-fix}
\bm\beta_k^{(i+1)}= \bm\beta_k^{(i)}   +       \frac{r_\theta}{100} \cdot  \mathrm{sign} (  \bm{\zeta}_k^{(i+1)} ),
\end{align}
where $r_\theta= \pi/\hat L$ stands for the grid interval, and $\mathrm{sign}(\cdot) $  stands for the signum function.

Following similar procedures to these in Appendix\ref{lemmaoff}, we can obtain the derivative of the objective function w.r.t $\bm\varphi_k$ as
\begin{align}
\bm{\varsigma}^{(i+1)}_k=[\varsigma^{(i+1)}(\varphi_{k,1}),\varsigma^{(i+1)}(\varphi_{k,2}),\ldots, \varsigma^{(i+1)}(\varphi_{k,\hat L})]^T, \label{eqderbeta-2D}
\end{align}
with
\begin{align}
&\varsigma^{(i+1)}(\varphi_{k,
l})=\notag\\
& 2 \mathrm{Re}\left(  (\a'_{\varphi} (\hat\vartheta_{ l}+ \beta_{k,l}^{(i)}, \varphi_{k,l}) )^H \X^H\X \a (\hat\vartheta_{ l}+ \beta_{k,l}^{(i)},\varphi_{k,l}) \right) \cdot c_{k1}^{(i+1)}\notag\\
&+ 2 \mathrm{Re}\left(  (\a'_{\varphi} (\hat\vartheta_{ l}+ \beta_{k,l}^{(i)}, \varphi_{k,l}))^H \X^H \c_{k2}^{(i+1)} \right), \label{eqderbeta1-2D}
\end{align}
where
$\a'_\varphi (\hat\vartheta_{l}+\beta_{ k,l}, \varphi_{k,l})= d \a(\hat\vartheta_{l}+\beta_{ k,l}, \varphi_{k,l}) /{d \varphi_{k,l}} $.
With (\ref{eqderbeta-2D}), we can update ${\bm\varphi}_k$ similarly to (\ref{equpv3}).
As mentioned in \cite{dai2018fdd}, the  elevation angle ranges from $-\pi/2$ to $\pi/2$, but it is sufficient to assume that $\varphi_{k,l}$ ranges from $0$ to $\pi/2$, because the steering vector contains $\cos\varphi_{k,l}$ only. Therefore, we initialize each $\varphi_{k,l}$ uniformly from $[0, \pi/2]$, and use a fixed stepsize to update ${\bm\varphi}_k$ [similarly to (\ref{equpv3-fix})]:
\begin{align}\label{equpv3-fix-2D}
{\bm\varphi}^{(i+1)}_k= {\bm\varphi}^{(i)}_k   +  \frac{\pi}{36} \cdot \max \left\{ (\varrho)^{i}, 0.001  \right\} \cdot \mathrm{sign} (\bm{\varsigma}^{(i+1)}_k),
\end{align}
where $0.9474<\varrho<1$ is a constant \cite{dai2018fdd}.

Once the algorithm converges, the estimated downlink channels $\h_k^e$s can be calculated as
\begin{align}
\h_k^e= \A_{\Omega_k}(\bm\beta_k,\bm\varphi_k) \left(\bm\Phi_{\Omega_k}(\bm\beta_k,\bm\varphi_k)\right)^{\dag} \y_k.
\end{align}

\begin{figure}
\center
\begin{tikzpicture}[scale=0.9]
\begin{semilogyaxis}[
ylabel={NMSE},grid=both, title={(a)},
legend style={at={(0.769,1.05),font=\footnotesize},
anchor=north,legend columns=1},xmin=40,xmax=70]
\addplot[mark=asterisk,red]  coordinates{
  (     40,     0.0421)
  (     45,     0.0275)
  (     50,     0.0200)
  (     55,     0.0161)
  (     60,     0.0134)
  (     65,     0.0115)
  (     70,     0.0103)
};
\addplot[mark=o,blue]  coordinates{
  (     40,     0.0447)
  (     45,     0.0298)
  (     50,     0.0217)
  (     55,     0.0164)
  (     60,     0.0136)
  (     65,     0.0116)
  (     70,     0.0103)
};
\addplot[mark=diamond]  coordinates{
  (     40,    0.6253)
  (     45,    0.3412)
  (     50,    0.1072)
  (     55,    0.0598)
  (     60,    0.0462)
  (     65,    0.0332)
  (     70,    0.0254)
};
\addplot[mark=diamond*,black] coordinates{
  (     40,      0.2562)
  (     45,      0.1747)
  (     50,      0.1282)
  (     55,      0.1028)
  (     60,      0.0838)
  (     65,      0.0726)
  (     70,      0.0634)
};
\addplot[mark=triangle] coordinates{
  (     40,    0.0685)        
  (     45,    0.0611)        
  (     50,    0.0531)        
  (     55,    0.0504)        
  (     60,    0.0501)        
  (     65,    0.0479)        
  (     70,    0.0449)        
};
\addplot[mark=square] coordinates{
  (     40,   0.1741)       
  (     45,   0.1417)       
  (     50,   0.1196)       
  (     55,   0.1022)       
  (     60,   0.0899)       
  (     65,   0.0720)       
  (     70,   0.0605)       
};
\addplot[mark=triangle*,black] coordinates{
  (     40,   0.1754)
  (     45,   0.1293)
  (     50,   0.0997)
  (     55,   0.0813)
  (     60,   0.0731)
  (     65,   0.0663)
  (     70,   0.0643)
};
\legend{Our method, Group-SBL, Common-SBL, Joint-OMP, Individual-off-grid, Individual-SBL, Individual-DFT}
\end{semilogyaxis}
\end{tikzpicture}
~
\begin{tikzpicture}[scale=0.9]
\begin{semilogyaxis}[xlabel={Number of  training pilot symbols},
ylabel={NMSE},grid=both, title={(b)},
legend style={at={(0.769,1.15),font=\footnotesize},
anchor=north,legend columns=1},xmin=40,xmax=70]
\addplot[mark=asterisk,red]  coordinates{
  (     30,     0.1676)
  (     35,     0.1011)
  (     40,     0.0601)
  (     45,     0.0411)
  (     50,     0.0340)
  (     55,     0.0293)
  (     60,     0.0258)
  (     65,     0.0222)
  (     70,     0.0195)
};
\addplot[mark=o,blue]  coordinates{
  (     30,      1.0419)
  (     35,      0.9606)
  (     40,      0.4215)
  (     45,      0.1951)
  (     50,      0.1197)
  (     55,      0.0907)
  (     60,      0.0668)
  (     65,      0.0527)
  (     70,      0.0419)
};
\addplot[mark=diamond]  coordinates{
  (     30,       1.1483)
  (     35,       0.9764)
  (     40,       1.0907)
  (     45,       0.8017)
  (     50,       0.4572)
  (     55,       0.2520)
  (     60,       0.1284)
  (     65,       0.0894)
  (     70,       0.0671)
};
\addplot[mark=diamond*,black] coordinates{
  (     30,       0.5450)
  (     35,       0.3866)
  (     40,       0.2647)
  (     45,       0.1794)
  (     50,       0.1388)
  (     55,       0.1053)
  (     60,       0.0863)
  (     65,       0.0708)
  (     70,       0.0637)
};
\addplot[mark=triangle] coordinates{
  (     30,      0.1832)
  (     35,      0.1183)
  (     40,      0.0694)
  (     45,      0.0581)
  (     50,      0.0513)
  (     55,      0.0527)
  (     60,      0.0524)
  (     65,      0.0519)
  (     70,      0.0459)
};
\addplot[mark=square] coordinates{
  (     30,      0.3193)
  (     35,      0.2335)
  (     40,      0.1759)
  (     45,      0.1388)
  (     50,      0.1198)
  (     55,      0.1043)
  (     60,      0.0890)
  (     65,      0.0732)
  (     70,      0.0610)
};
\addplot[mark=triangle*,black] coordinates{
  (     30,      0.5472)
  (     35,      0.2717)
  (     40,      0.1792)
  (     45,      0.1257)
  (     50,      0.0996)
  (     55,      0.0827)
  (     60,      0.0717)
  (     65,      0.0658)
  (     70,      0.0644)
};
\end{semilogyaxis}
\end{tikzpicture}
\caption{NMSE of downlink channel estimate versus the number of  training pilot symbols for ULA, where $N=80$, $K=60$, $G=3$ and SNR$=0$ dB. a) $L_s=4$ and $L_v=0$; b) $L_s=2$ and $L_v=2$.
 }\label{fig-vsT}
\end{figure}
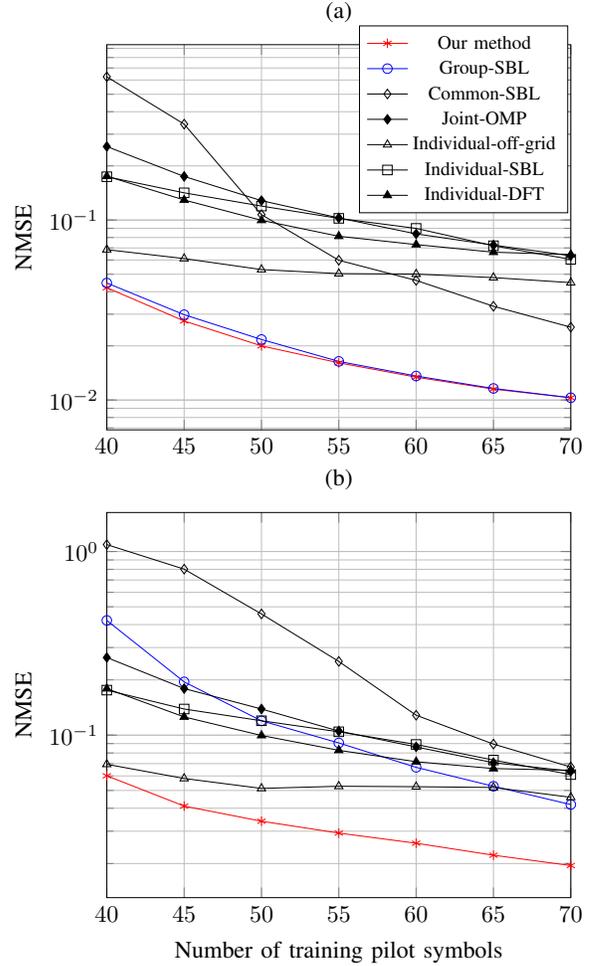

\section{Simulation Results}

In this section, numerical simulations are conducted to evaluate the performance of our proposed method.
The proposed method is compared with the following baselines:
\begin{itemize}
\item \textbf{Baseline~1} (Individual-DFT): Each $\h_k$ is individually recovered using the  $l_1$-norm minimization algorithm \cite{donoho2006compressed,candes2006robust} with a DFT basis.

\item \textbf{Baseline~2} (Individual-SBL): Each $\h_k$ is individually recovered using the standard SBL method \cite{tipping2001sparse} with a DFT basis.

\item \textbf{Baseline~3} (Individual-off-grid): Each $\h_k$ is individually recovered using the off-grid SBL method \cite{dai2018fdd}.


\item \textbf{Baseline~4} (Joint-OMP): $\h_k$s are jointly recovered using  the joint
orthogonal matching pursuit recovery method \cite{rao2014distributed}.

\item \textbf{Baseline~5} (Common-SBL):  $\h_k$s are jointly recovered using the multiple measurement SBL method \cite{yang2013off} with an off-grid basis, where $\h_k$s are assumed to share a uniform sparsity structure.

\item \textbf{Baseline~6} (Group-SBL):  $\h_k$s are jointly recovered using the group SBL method \cite{wang2016novel}  with an off-grid basis.\footnote{For fairness, the off-grid refinement method used in \cite{wang2016novel} is replaced by the one used in ours.}

\end{itemize}

We first focus on simulations for ULAs, where we use the 3GPP spatial channel model (SCM) \cite{molisch2003geometry} to generate the channels  for an urban microcell.
The downlink frequency is $2170$~MHz and the inter-antenna spacing is $d=c/(2f_0)$, with $c$ being the light speed and $f_0=2000$~MHz.  Then, we run simulations with the 3GPP 3D channel model \cite{3gpp-3D-Model}, which provides a 2D array model. All the parameters of the 3D channel model follow 3D-UMa-NOLS (see Table 7.3-6 in \cite{3gpp-3D-Model}).
The normalized mean square error (NMSE) is defined as
\begin{align}
\frac{1}{M_c}\sum_{m=1}^{M_c} \frac{  \sum_{k=1}^K   \| \tilde{\h}_k^m - \h_k^m  \|_2^2      }{    \sum_{k=1}^K \| \h_k^m  \|_2^2},
\end{align}
where $\h_k^m$ is the downlink channel vector for the $k$-th MU at the $m$-th Monte Carlo trial,
$\tilde{\h}_k^m$ is the estimate of $ \h_k^m $,  and $M_c=200$ is the number of Monte Carlo trials.
Unless otherwise specified, in the following, we assume that every  channel realization consists of $N_c$ random scattering clusters, each cluster contains $N_s=20$ sub-paths concentrated in a $\mathcal{A}=10^\circ$ angular spread, and the number of grid points is fixed at $\hat L=N$.

\subsection{Channel Estimation Performance Versus $T$ for ULA}

In Fig.~\ref{fig-vsT},  Monte Carlo trials are carried out to investigate the impact of the number of pilot symbols on the channel estimation performance for ULA. Assume that a ULA is equipped at the BS with $N=80$ antennas and the system supports $K=60$ MUs. The MUs are randomly dropped into three groups with a uniform distribution. The number of shared (unshared) scattering clusters for users in the same group is denoted by $L_s$ ($L_v$).\footnote{Note that $L_s+L_v=N_c$, and two users sharing a scattering cluster means that the AoD mean of the scattering cluster is the same.} If $L_s=N_c$, it means that users in the same group have a uniform scattering structure, while if $L_s=0$, there is no group property for users.
The center AoD of each scattering clusters ranges from $-90^\circ$ to $90^\circ$ uniformly.
The training pilots are randomly generated, and the SNR is chosen as  $0$ dB.
Fig.~\ref{fig-vsT} shows the NMSE performance of the downlink channel estimate achieved by the different channel estimation strategies versus the number of training pilot symbols $T$. All the results are obtained by averaging over $200$ Monte Carlo channel realizations.
It can be seen that 1) the NMSEs of all the methods decrease as the number of training pilot symbols increases;
2) compared with the individual recovery methods (Individual-DFT, Individual-SBL, Individual-off-grid), Joint-OMP and Common-SBL, our method and  Group-SBL can improve the NMSE performance due to exploiting the common sparsity among nearby users;
3) when the uniform shared sparsity assumption holds true for each group ($L_s=4$ and $L_v=0$), our method and Group-SBL achieve similar channel estimation performance (Fig.~\ref{fig-vsT}-a), which verifies that removing the DP prior in our method does not bring any performance loss;
and 4) when the uniform sparsity assumption fails to hold ($L_s=2$ and $L_v=2$), our method outperforms Group-SBL because our method can handle outliers but Group-SBL is only designed for the uniform sparsity assumption.

\subsection{Channel Estimation Performance Versus SNR for ULA}

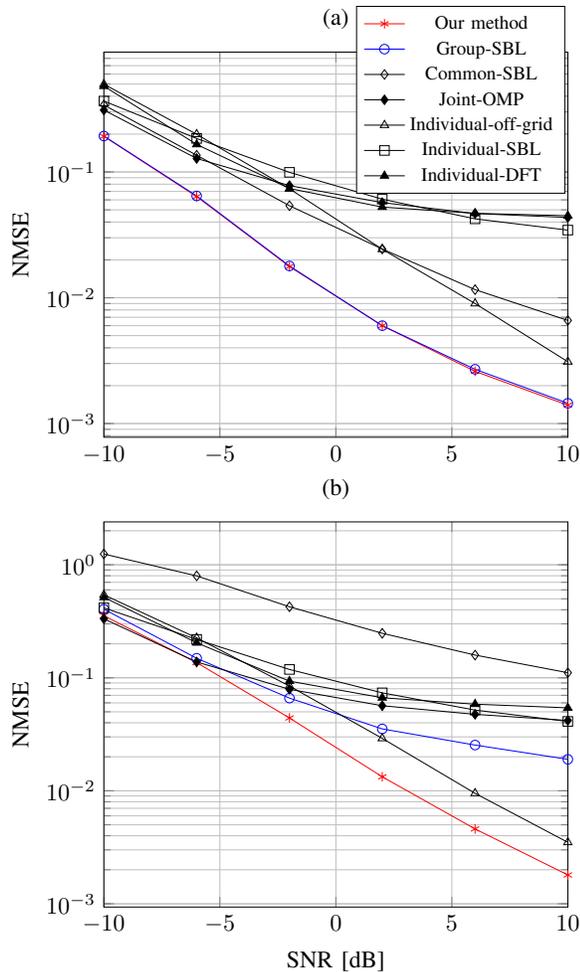
\begin{figure}
\center
\begin{tikzpicture}[scale=0.9]
\begin{semilogyaxis}[
ylabel={NMSE},grid=both, title={(a)},
legend style={at={(0.769,1.12),font=\footnotesize},
anchor=north,legend columns=1},xmin=-10,xmax=10]
\addplot[mark=asterisk,red]  coordinates{
  (    -10 ,      0.1920)
  (     -6 ,      0.0637)
  (     -2 ,      0.0177)
  (      2 ,      0.0060)
  (      6 ,      0.0026)
  (     10 ,      0.0014)
};
\addplot[mark=o,blue]  coordinates{
  (    -10 ,     0.1931)
  (     -6 ,     0.0645)
  (     -2 ,     0.0179)
  (      2 ,     0.0060)
  (      6 ,     0.0027)
  (     10 ,     0.00145)
};
\addplot[mark=diamond]  coordinates{
  (    -10 ,      0.3407)
  (     -6 ,      0.1358)
  (     -2 ,      0.0539)
  (      2 ,      0.0244)
  (      6 ,      0.0116)
  (     10 ,      0.0066)
};
\addplot[mark=diamond*]  coordinates{
  (    -10 ,        0.3105)
  (     -6 ,        0.1280)
  (     -2 ,        0.0779)
  (      2 ,        0.0571)
  (      6 ,        0.0467)
  (     10 ,        0.0435)
};
\addplot[mark=triangle] coordinates{
  (    -10 ,      0.4984)    
  (     -6 ,      0.1993)    
  (     -2 ,      0.0735)    
  (      2 ,      0.0243)    
  (      6 ,      0.0090)    
  (     10 ,      0.0031)    
};
\addplot[mark=square] coordinates{
  (    -10 ,      0.3661)
  (     -6 ,      0.1847)
  (     -2 ,      0.0993)
  (      2 ,      0.0608)
  (      6 ,      0.0424)
  (     10 ,      0.0345)
};
\addplot[mark=triangle*,black] coordinates{
  (    -10 ,     0.4810)
  (     -6 ,     0.1656)
  (     -2 ,     0.0739)
  (      2 ,     0.0526)
  (      6 ,     0.0469)
  (     10 ,     0.0449)
};
\legend{Our method, Group-SBL, Common-SBL, Joint-OMP, Individual-off-grid, Individual-SBL, Individual-DFT}
\end{semilogyaxis}
\end{tikzpicture}
~
\begin{tikzpicture}[scale=0.9]
\begin{semilogyaxis}[xlabel={SNR [dB]},
ylabel={NMSE},grid=both, title={ (b)},
legend style={at={(0.769,1.15),font=\footnotesize},
anchor=north,legend columns=1},xmin=-10,xmax=10]
\addplot[mark=asterisk,red]  coordinates{
  (    -10 ,      0.3540)
  (     -6 ,      0.1363)
  (     -2 ,      0.0441)
  (      2 ,      0.0133)
  (      6 ,      0.0046)
  (     10 ,      0.0018)
};
\addplot[mark=o,blue]  coordinates{
  (    -10 ,       0.4045)
  (     -6 ,       0.1489)
  (     -2 ,       0.0661)
  (      2 ,       0.0353)
  (      6 ,       0.0254)
  (     10 ,       0.0190)
};
\addplot[mark=diamond]  coordinates{
  (    -10 ,        1.2458)
  (     -6 ,        0.7976)
  (     -2 ,        0.4250)
  (      2 ,        0.2476)
  (      6 ,        0.1589)
  (     10 ,        0.1109)
};
\addplot[mark=diamond*]  coordinates{
  (    -10 ,       0.3326)
  (     -6 ,       0.1381)
  (     -2 ,       0.0791)
  (      2 ,       0.0566)
  (      6 ,       0.0475)
  (     10 ,       0.0419)
};
\addplot[mark=triangle] coordinates{
  (    -10 ,       0.5420)
  (     -6 ,       0.2256)
  (     -2 ,       0.0845)
  (      2 ,       0.0291)
  (      6 ,       0.0095)
  (     10 ,       0.0035)
};
\addplot[mark=square] coordinates{
  (    -10 ,       0.4156)
  (     -6 ,       0.2185)
  (     -2 ,       0.1185)
  (      2 ,       0.0736)
  (      6 ,       0.0518)
  (     10 ,       0.0411)
};
\addplot[mark=triangle*,black] coordinates{
  (    -10 ,       0.5155)
  (     -6 ,       0.2040)
  (     -2 ,       0.0934)
  (      2 ,       0.0666)
  (      6 ,       0.0584)
  (     10 ,       0.0542)
};
\end{semilogyaxis}
\end{tikzpicture}
\caption{NMSE of downlink channel estimate versus SNR for ULA, where $N=80$, $G=4$, $K=50$ and $T=60$. a) $L_s=3$ and $L_v=0$; b) $L_s=2$ and $L_v=1$.
 }\label{fig-vsSNR}
\end{figure}

In Fig.~\ref{fig-vsSNR}, we study the impact of SNR on the channel estimation performance for ULA. We consider the same scenario as in Section~V-A, except that the number of training pilot symbols is fixed at $60$ and the number of users is set to $50$.
Fig.~\ref{fig-vsSNR} shows the NMSE performance of the downlink channel estimate achieved
by the different channel estimation strategies versus SNR.
All the results are obtained by averaging over 200 Monte Carlo channel realizations.
It is shown that 1) the NMSEs of all the methods decrease as SNR increases;
2) when the uniform shared sparsity assumption holds true, our method and  Group-off-grid  achieve very similar channel estimation performance (Fig.~\ref{fig-vsSNR}-a); 3) when the uniform sparsity assumption fails to hold,  Group-off-grid gives very bad performance because of outliers deviated from the group sparsity patterns (Fig.~\ref{fig-vsSNR}-b); and 4) the proposed general sparsity model can  capture the true group sparse structure, and our method indeed works for the general sparsity model and  can significantly improve the channel estimation performance.


\begin{figure}
\center
\begin{tikzpicture}[scale=0.9]
\begin{axis}[xlabel={SNR [dB]},
ylabel={Sum Rate [bit/s/Hz]},grid=both, 
legend style={at={(0.239,1.06),font=\footnotesize},
anchor=north,legend columns=1},xmin=-10,xmax=10]
\addplot[mark=asterisk,red]  coordinates{
  (    -10 ,     19.7931)
  (     -6 ,     29.5466)
  (     -2 ,     38.7470)
  (      2 ,     48.7885)
  (      6 ,     57.4904)
  (     10 ,     66.0794)
};
\addplot[mark=o,blue]  coordinates{
  (    -10 ,     18.5878)
  (     -6 ,     27.7674)
  (     -2 ,     35.2888)
  (      2 ,     42.6726)
  (      6 ,     49.9982)
  (     10 ,     56.5844)
};
\addplot[mark=diamond]  coordinates{
  (    -10 ,       5.8048)
  (     -6 ,      12.5672)
  (     -2 ,      17.5605)
  (      2 ,      22.7137)
  (      6 ,      24.6388)
  (     10 ,      25.7426)
};
\addplot[mark=diamond*]  coordinates{
  (    -10 ,      9.2089)
  (     -6 ,     13.5800)
  (     -2 ,     19.7066)
  (      2 ,     25.5708)
  (      6 ,     28.0805)
  (     10 ,     29.9783)
};
\addplot[mark=triangle] coordinates{
  (    -10 ,        7.3608)
  (     -6 ,       13.5465)
  (     -2 ,       21.1846)
  (      2 ,       26.6198)
  (      6 ,       30.1861)
  (     10 ,       31.3374)
};
\addplot[mark=square] coordinates{
  (    -10 ,      7.7105)
  (     -6 ,     13.9541)
  (     -2 ,     19.5450)
  (      2 ,     24.1376)
  (      6 ,     27.7051)
  (     10 ,     28.0875)
};
\addplot[mark=triangle*,black] coordinates{
  (    -10 ,      7.5321)
  (     -6 ,     14.0419)
  (     -2 ,     19.6444)
  (      2 ,     24.0515)
  (      6 ,     26.3949)
  (     10 ,     27.0564)
};
\legend{Our method, Group-SBL, Common-SBL, Joint-OMP, Individual-off-grid, Individual-SBL, Individual-DFT}
\end{axis}
\end{tikzpicture}
\caption{Sum spectral efficiency versus SNR for JSDM, where $N=80$, $G=4$, $K=100$, $T=60$, $L_s=2$ and $L_v=1$.
 }\label{fig-vsSNR-Sumrate}
\end{figure}
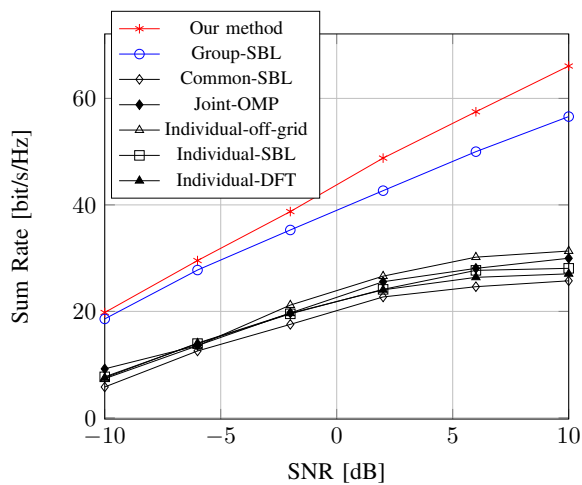

\begin{figure}
\center
\begin{tikzpicture}[scale=0.9]
\begin{axis}[xlabel={Angular spread [degree]},
ylabel={Sum Rate [bit/s/Hz]},grid=both, 
legend style={at={(0.769,1.15),font=\footnotesize},
anchor=north,legend columns=1},xmin=4,xmax=20]
\addplot[mark=asterisk,red]  coordinates{
  (    4,    51.2475)
  (    8,    49.0277)
  (   12,    41.9924)
  (   16,    36.1741)
  (   20,   33.7687)
};
\addplot[mark=o,blue]  coordinates{
  (    4,    47.6872)
  (    8,    43.6923)
  (   12,    37.1285)
  (   16,    28.8685)
  (   20,    27.0614)
};
\addplot[mark=diamond]  coordinates{
  (    4,    27.1413)
  (    8,    18.0549)
  (   12,   14.0636)
  (   16,   11.6587)
  (   20,    10.9608)
};
\addplot[mark=diamond*]  coordinates{
  (     4,    30.7199)
  (     8,    22.2816)
  (    12,    19.5336)
  (    16,    18.9588)
  (    20,    18.2048)
};
\addplot[mark=triangle] coordinates{
  (    4,   30.7950)
  (    8,   24.6571)
  (   12,   22.5506)
  (   16,   21.3658)
  (   20,   21.0129)
};
\addplot[mark=square] coordinates{
  (    4,   28.3145)
  (    8,   22.5282)
  (   12,   21.0013)
  (   16,   19.6771)
  (   20,   18.6878)
};
\addplot[mark=triangle*,black] coordinates{
  (   4,   29.1060)
  (   8,   23.1493)
  (  12,   20.9555)
  (  16,   19.9831)
  (  20,   18.6191)
};
\legend{Our method, Group-SBL, Common-SBL, Joint-OMP, Individual-off-grid, Individual-SBL, Individual-DFT}
\end{axis}
\end{tikzpicture}
\caption{Sum spectral efficiency versus the angular spread for JSDM, where $N=80$, $G=4$, $K=100$, $L_s=2$, $L_v=1$, $T=50$ and  SNR$=0$ dB.
}\label{fig-vsA-Sumrate}
\end{figure}
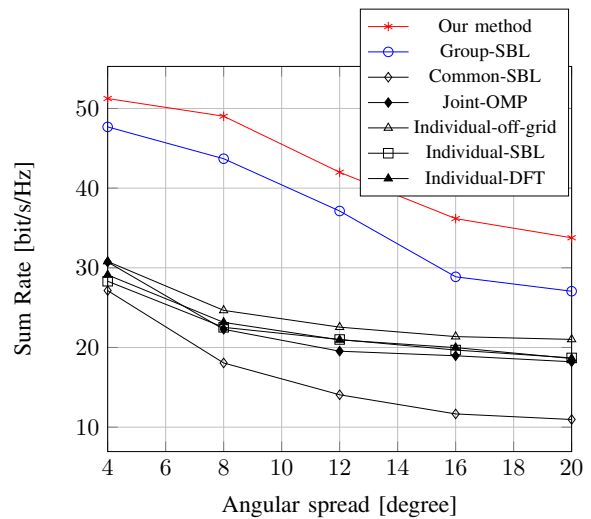

\subsection{Comparison of Sum Spectral Efficiency for JSDM}
In Figs.~\ref{fig-vsSNR-Sumrate} and \ref{fig-vsA-Sumrate}, we study the sum spectral efficiency when the proposed method is integrated into JSDM.  Assume that a ULA is equipped at BS with $N=80$ antennas and the system supports $K=100$ MUs. The MUs are randomly dropped into four groups with a uniform distribution, and $20\%$ of MUs will be activated in the system.
Following \cite{nam2014joint}, the standard $K$-Means algorithm is chosen to cluster users if the method (e.g., Individual-DFT, Individual-SBL, Individual-off-grid, Joint-OMP and Common-SBL) cannot provide knowledge of the user grouping, and then zero-forcing beamforming (ZFBF) with semi-orthogonal user selection (SUS) is adopted for each group in the JSDM framework.
Fig.~\ref{fig-vsSNR-Sumrate} shows the  sum spectral efficiency achieved
by the different strategies versus SNR, and Fig.~\ref{fig-vsA-Sumrate} shows the  sum spectral efficiency achieved
by the different strategies versus angular spread.
All the results are obtained by averaging over 200 Monte Carlo channel realizations.
Compared with other methods, ours can significantly improve the sum-rate performance of JSDM systems. This is because our method can give better channel estimation and user grouping results in the sense of Bayesian optimality, so as to alleviate the interference across different groups.

\subsection{Channel Estimation Performance Versus Inexact $G$}
In Fig.~\ref{fig-vsGinexact}, we illustrate that our method applies to unknown real number of user groups $G^\star$.
Assume that a ULA is equipped at the BS with $N=100$ antennas and  the system supports $K=50$ MUs. The number of training pilot symbols
is fixed at 60, $L_s=2$ and  $L_v=1$. The MUs are randomly dropped into four groups
with a uniform distribution, but the real number of user groups is not exactly known. Fig.~\ref{fig-vsGinexact} shows the NMSE performance of the downlink channel estimate achieved by the different channel estimation strategies versus  an inexact $G$.
It is interesting to see that most curves in the figure remain unchanged.
The reason that NMSEs do not change much for the individual methods (Individual-DFT, Individual-SBL, Individual-off-grid) is  because each $\h_k$ is estimated individually  for each user, and  thus its estimation performance is not related to $G$ or $G^\star$, while Common-SBL and Joint-OMP always assume that there is just one group ($G=1$).
The reason why the NMSE of our method also does not change much is  because the adopted general model can capture a much more general group sparse structure and can provide a robust result for an inexact choice of $G$, as long as $G$ is not much smaller than the true value $G^\star$. Hence, Fig.~\ref{fig-vsGinexact} verifies that our method works well for an unknown $G$.

\subsection{Channel Estimation and Sum Rate Performance with 2D Array}
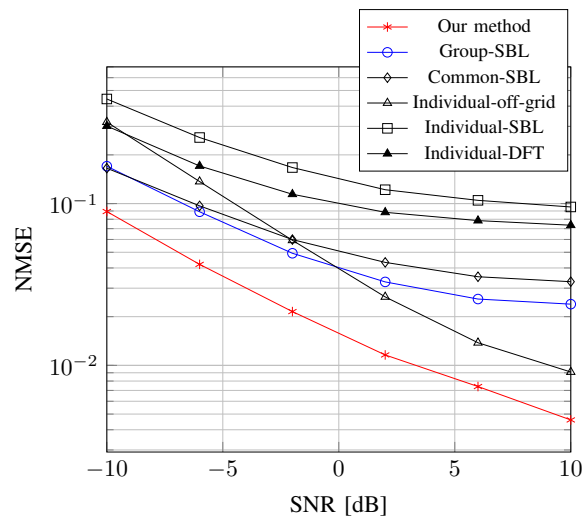
\begin{figure}
\center
\begin{tikzpicture}[scale=0.9]
\begin{semilogyaxis}[xlabel={SNR [dB]},
ylabel={NMSE},grid=both, 
legend style={at={(0.769,1.15),font=\footnotesize},
anchor=north,legend columns=1},xmin=-10,xmax=10]
\addplot[mark=asterisk,red]  coordinates{
  (    -10 ,      0.0893)
  (     -6 ,      0.0421)
  (     -2 ,      0.0215)
  (      2 ,      0.0116)
  (      6 ,      0.0074)
  (     10 ,      0.0046)
};
\addplot[mark=o,blue]  coordinates{
  (    -10 ,      0.1702)
  (     -6 ,      0.0890)
  (     -2 ,      0.0493)
  (      2 ,      0.0328)
  (      6 ,      0.0257)
  (     10 ,      0.0239)
};
\addplot[mark=diamond]  coordinates{
  (    -10 ,      0.1655)
  (     -6 ,      0.0968)
  (     -2 ,      0.0599)
  (      2 ,      0.0433)
  (      6 ,      0.0353)
  (     10 ,      0.0329)
};
\addplot[mark=triangle] coordinates{
  (    -10 ,       0.3190)    
  (     -6 ,       0.1368)    
  (     -2 ,       0.0593)    
  (      2 ,       0.0265)    
  (      6 ,       0.0138)    
  (     10 ,       0.0091)    
};
\addplot[mark=square] coordinates{
  (    -10 ,       0.4431)
  (     -6 ,       0.2560)
  (     -2 ,       0.1670)
  (      2 ,       0.1218)
  (      6 ,       0.1047)
  (     10 ,       0.0953)
};
\addplot[mark=triangle*,black] coordinates{
  (    -10 ,       0.3015)
  (     -6 ,       0.1708)
  (     -2 ,       0.1142)
  (      2 ,       0.0883)
  (      6 ,       0.0785)
  (     10 ,       0.0735)
};
\legend{Our method, Group-SBL, Common-SBL,  Individual-off-grid, Individual-SBL, Individual-DFT}
\end{semilogyaxis}
\end{tikzpicture}
\caption{NMSE of downlink channel estimate versus SNR for 2D array, where $N=10\times 10$, $G=4$, $K=30$, $T=60$, $L_s=2$ and $L_v=1$.
 }\label{fig-vsSNR3D-NMSE}
\end{figure}

\begin{figure}
\center
\begin{tikzpicture}[scale=0.9]
\begin{axis}[xlabel={SNR [dB]},
ylabel={Sum Rate [bit/s/Hz]},grid=both, 
legend style={at={(0.239,0.99),font=\footnotesize},
anchor=north,legend columns=1},xmin=-10,xmax=10]
\addplot[mark=asterisk,red]  coordinates{
  (    -10 ,     21.9572)
  (     -6 ,     28.2848)
  (     -2 ,     33.8521)
  (      2 ,     39.2468)
  (      6 ,     44.3493)
  (     10 ,     49.3793)
};
\addplot[mark=o,blue]  coordinates{
  (    -10 ,      18.5191)
  (     -6 ,      23.9628)
  (     -2 ,      28.2782)
  (      2 ,      31.7685)
  (      6 ,      33.4141)
  (     10 ,      34.6278)
};
\addplot[mark=diamond]  coordinates{
  (    -10 ,        18.3301)
  (     -6 ,        22.9958)
  (     -2 ,        26.2663)
  (      2 ,        28.7191)
  (      6 ,        29.6332)
  (     10 ,        30.3902)
};
\addplot[mark=triangle] coordinates{
  (    -10 ,       16.7478)
  (     -6 ,       23.1922)
  (     -2 ,       29.2046)
  (      2 ,       35.6216)
  (      6 ,       41.0305)
  (     10 ,       45.7091)
};
\addplot[mark=square] coordinates{
  (    -10 ,       15.1483)
  (     -6 ,       20.1402)
  (     -2 ,       24.0868)
  (      2 ,       27.9331)
  (      6 ,       30.6352)
  (     10 ,       32.6314)
};
\addplot[mark=triangle*,black] coordinates{
  (    -10 ,      16.6936)
  (     -6 ,      21.9583)
  (     -2 ,      25.8222)
  (      2 ,      29.3234)
  (      6 ,      31.4531)
  (     10 ,      33.1365)
};
\legend{Our method, Group-SBL, Common-SBL, Individual-off-grid, Individual-SBL, Individual-DFT}
\end{axis}
\end{tikzpicture}
\caption{Sum spectral efficiency versus SNR for 2D array, where $N=10\times 10$, $G=4$, $K=30$, $T=60$, $L_s=2$ and $L_v=1$.
 }\label{fig-vsSNR3D-sumrate}
\end{figure}
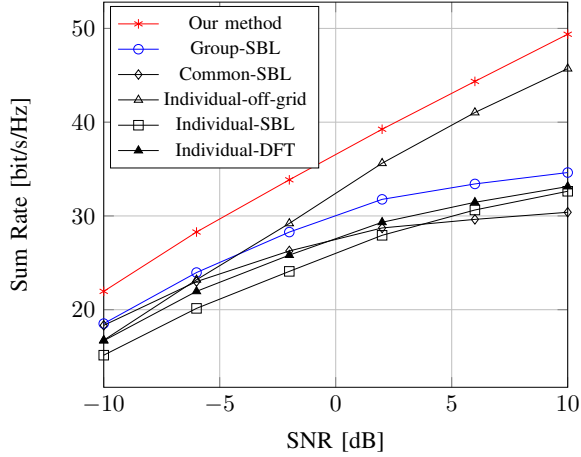

In Figs.~\ref{fig-vsT3D-NMSE}--\ref{fig-vsSNR3D-sumrate}, Monte Carlo trials are carried out to investigate the channel estimation and sum rate performance with the 2D array. Assume that the 2D planar array at the BS is equipped with $10\times 10$ antennas, where
both the horizontal and vertical inter-antenna spacings are a half wavelength.
Every channel realization consists of $N_c=3$ random scattering clusters (with $L_s=2$ and $L_v=1$), and each cluster contains $N_s=20$ subpaths.
The AoDs are randomly generated in the 3GPP 3D channel model, where the azimuth AoDs range from $-180^\circ$ to $180^\circ$ and the elevation AoDs range from $-90^\circ$ to $90^\circ$.
The system supports $K=30$ MUs simultaneously, and  minimum mean-squared error (MMSE) precoder is adopted at the BS.
All the results are obtained by averaging over $200$ Monte Carlo channel realizations.
Figs.~\ref{fig-vsT3D-NMSE} and \ref{fig-vsT3D-sumrate} show the NMSE and the sum spectral
efficiency achieved by the different strategies versus the number of training pilot symbols $T$, respectively, and
Figs.~\ref{fig-vsSNR3D-NMSE} and \ref{fig-vsSNR3D-sumrate} show the NMSE and the sum spectral
efficiency achieved by the different strategies versus SNR, respectively.
It can be seen that our proposed  method indeed works for the 2D array, and the results reverify that our method can substantially improve the channel estimation performance, as well as the sum spectral efficiency.


\section{Conclusion}
The problem  of joint downlink channel estimation and user grouping in massive MIMO systems is addressed
in this paper. We first provide a general model to capture a more general sparse structure for user grouping. Then, we propose an SBL-based framework to handle the general sparsity model, which can fully exploit the common sparsity to cluster nearby users and exclude the harmful effect from outliers simultaneously.
To the best of our knowledge, our
work is the first to utilize an off-grid SBL-based framework
to jointly estimate the channel and cluster the users. Simulation results demonstrate that our method indeed works for the general sparsity model and can significantly improve the channel estimation performance when the uniform sparsity assumption fails to hold. Moreover, it is worth noting that extending our method with the DP prior and an automatically determined $G$ is straightforward.

\begin{figure}
\center
\begin{tikzpicture}[scale=0.9]
\begin{semilogyaxis}[xlabel={$G$},
ylabel={NMSE},grid=both, 
legend style={at={(0.769,1.15),font=\footnotesize},
anchor=north,legend columns=1},xmin=2,xmax=10]
\addplot[mark=asterisk,red]  coordinates{
  (    1,     0.0178)
  (    2,     0.0168)
  (    4,     0.0161)
  (    6,     0.0154)
  (    8,     0.0151)
  (   10,     0.0150)
};
\addplot[mark=o,blue]  coordinates{
  (    1,      0.1899)
  (    2,      0.0635)
  (    4,      0.0359)
  (    6,      0.0270)
  (    8,      0.0225)
  (   10,      0.0218)
};
\addplot[mark=diamond]  coordinates{
  (    1,      0.1775)
  (    2,      0.1871)
  (    4,      0.1932)
  (    6,      0.1847)
  (    8,      0.1752)
  (   10,      0.2026)
};
\addplot[mark=diamond*]  coordinates{
  (    1,      0.0769)
  (    2,      0.0779)
  (    4,      0.0776)
  (    6,      0.0775)
  (    8,      0.0770)
  (   10,      0.0761)
};
\addplot[mark=triangle] coordinates{
  (    1,     0.0249)
  (    2,     0.0247)
  (    4,     0.0248)
  (    6,     0.0249)
  (    8,     0.0246)
  (   10,     0.0250)
};
\addplot[mark=square] coordinates{
  (    1,       0.0852)
  (    2,       0.0858)
  (    4,       0.0851)
  (    6,       0.0844)
  (    8,       0.0841)
  (   10,       0.0852)
};
\addplot[mark=triangle*,black] coordinates{
  (    1,      0.0664)
  (    2,      0.0664)
  (    4,      0.0659)
  (    6,      0.0654)
  (    8,      0.0649)
  (   10,      0.0661)
};
\legend{Our method, Group-SBL, Common-SBL, Joint-OMP, Individual-off-grid, Individual-SBL, Individual-DFT}
\end{semilogyaxis}
\end{tikzpicture}
\caption{NMSE of downlink channel estimate versus inexact $G$, where $G^\star=4$, $N=100$, $K=50$, $T=60$, $L_s=2$, $L_v=1$, and SNR$=0$ dB.
 }\label{fig-vsGinexact}
\end{figure}
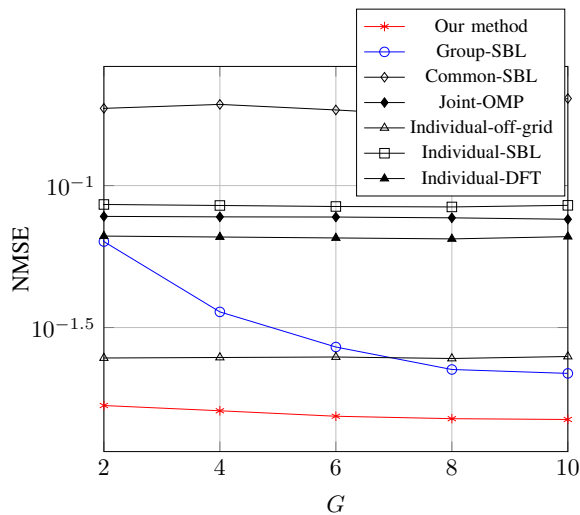

\begin{figure}
\center
\begin{tikzpicture}[scale=0.9]
\begin{semilogyaxis}[xlabel={Number of  training pilot symbols},
ylabel={NMSE},grid=both, 
legend style={at={(0.769,1.16),font=\footnotesize},
anchor=north,legend columns=1},xmin=40,xmax=70]
\addplot[mark=asterisk,red]  coordinates{
  (     40,     0.0326)
  (     45,     0.0265)
  (     50,     0.0223)
  (     55,     0.0188)
  (     60,     0.0161)
  (     65,     0.0140)
  (     70,     0.0121)
};
\addplot[mark=o,blue]  coordinates{
  (     40,      0.1105)
  (     45,      0.0805)
  (     50,      0.0611)
  (     55,      0.0491)
  (     60,      0.0407)
  (     65,      0.0349)
  (     70,      0.0296)
};
\addplot[mark=diamond]  coordinates{
  (     40,       0.1239)
  (     45,       0.0975)
  (     50,       0.0787)
  (     55,       0.0637)
  (     60,       0.0514)
  (     65,       0.0426)
  (     70,       0.0353)
};
\addplot[mark=triangle] coordinates{
  (     40,      0.0665)
  (     45,      0.0571)
  (     50,      0.0484)
  (     55,      0.0413)
  (     60,      0.0345)
  (     65,      0.0280)
  (     70,      0.0236)
};
\addplot[mark=square] coordinates{
  (     40,     0.2531)
  (     45,     0.2174)
  (     50,     0.1910)
  (     55,     0.1635)
  (     60,     0.1413)
  (     65,     0.1230)
  (     70,     0.1063)
};
\addplot[mark=triangle*,black] coordinates{
  (     40,      0.1948)
  (     45,      0.1639)
  (     50,      0.1395)
  (     55,      0.1176)
  (     60,      0.1008)
  (     65,      0.0884)
  (     70,      0.0776)
};
\legend{Our method, Group-SBL, Common-SBL, Individual-off-grid, Individual-SBL, Individual-DFT}
\end{semilogyaxis}
\end{tikzpicture}
\caption{NMSE of downlink channel estimate versus $T$ for 2D array, where $N=10\times10$, $G=4$, $K=30$, SNR$=0$ dB, $L_s=2$ and $L_v=1$.
 }\label{fig-vsT3D-NMSE}
\end{figure}

\begin{figure}
\center
\begin{tikzpicture}[scale=0.9]
\begin{axis}[xlabel={Number of  training pilot symbols},
ylabel={Sum Rate [bit/s/Hz]},grid=both, 
legend style={at={(0.239,1.05),font=\footnotesize},
anchor=north,legend columns=1},xmin=40,xmax=70,ymax=45]
\addplot[mark=asterisk,red]  coordinates{
  (     40,      32.0114)
  (     45,      33.6274)
  (     50,      34.8804)
  (     55,      35.9438)
  (     60,      36.7407)
  (     65,      37.4618)
  (     70,      37.8033)
};
\addplot[mark=o,blue]  coordinates{
  (     40,      20.9766)
  (     45,      25.4810)
  (     50,      27.5727)
  (     55,      29.2773)
  (     60,      30.4093)
  (     65,      31.2957)
  (     70,      32.1218)
};
\addplot[mark=diamond]  coordinates{
  (     40,      20.5501)
  (     45,      22.7659)
  (     50,      24.7622)
  (     55,      26.4613)
  (     60,      28.0401)
  (     65,      29.2867)
  (     70,      30.3590)
};
\addplot[mark=triangle] coordinates{
  (     40,      30.5810)
  (     45,      31.3186)
  (     50,      31.8987)
  (     55,      32.5013)
  (     60,      33.1355)
  (     65,      33.6468)
  (     70,      34.3945)
};
\addplot[mark=square] coordinates{
  (     40,      21.8387)
  (     45,      23.2780)
  (     50,      24.5472)
  (     55,      25.6813)
  (     60,      26.6422)
  (     65,      27.3934)
  (     70,      28.2201)
};
\addplot[mark=triangle*,black] coordinates{
  (     40,      23.1263)
  (     45,      24.6886)
  (     50,      26.0012)
  (     55,      27.0756)
  (     60,      28.0234)
  (     65,      28.6743)
  (     70,      29.3279)
};
\legend{Our method, Group-SBL, Common-SBL, Individual-off-grid, Individual-SBL, Individual-DFT}
\end{axis}
\end{tikzpicture}
\caption{Sum spectral efficiency versus $T$ for 2D array, where $N=10\times10$, $G=4$, $K=30$, SNR$=0$ dB, $L_s=2$ and $L_v=1$.
 }\label{fig-vsT3D-sumrate}
\end{figure}


\newpage

\section*{Appendix}

\appendices

\subsection{Proof of Lemma~3}\label{lemmaa}

The objective function  in (\ref{eqM1}) can be rewritten as
\begin{align}
&\mathcal{U}(  q_{1}, q_{2}^{(i)},q_{3}^{(i)},q_{4}^{(i)},q_{5}^{(i)}\notag\\
=&\int  q_{1}q_{2}^{(i)}q_{3}^{(i)}q_{4}^{(i)}q_{5}^{(i)}  \ln \frac{ p(\Y,\bm\Theta)}{ q_{1} q_{2}^{(i)}q_{3}^{(i)}q_{4}^{(i)}q_{5}^{(i)} } d\bm\Theta)\label{eq-dq1}\\
\propto&\int  q_{1}  \left<   \ln p( \Y, \bm\Theta )   \right>_{q^{(i)}(\bm \Theta \setminus \Theta_1)}  d \Theta_1
- \int  q_{1} \ln q_{1} d \Theta_{1}\\
=& \int  q_{1} \ln \frac{ \exp \left(\left<   \ln p( \Y, \bm\Theta )   \right>_{q^{(i)}(\bm \Theta \setminus \Theta_1)}    \right  )}{ q_{1} } d\Theta_1  \\
\le& \ln \int  q_{1} \frac{ \exp \left(\left<   \ln p( \Y, \bm\Theta )   \right>_{q^{(i)}(\bm \Theta \setminus \Theta_1)}    \right  )}{ q_{1} } d\Theta_1\label{eqjason}\\
=& \ln \int \exp \left(\left<   \ln p( \Y, \bm\Theta )   \right>_{q^{(i)}(\bm \Theta \setminus \Theta_1)}    \right  ) d\Theta_1,\label{eq-dq16}
\end{align}
where $\bm\Theta \setminus \Theta_j$ stands for the set $\bm\Theta$ excluding $\Theta_j$, and Jensen's inequality is applied in (\ref{eqjason}). Clearly, the objective function in  (\ref{eqM1}) is maximized, if the inequality in  (\ref{eqjason}) holds strictly, which means  the optimization problem (\ref{eqM1}) has a unique solution:
\begin{align}
&\ln q^{(i+1)}(\alpha)\notag\\
\propto&   \left<   \ln p( \Y, \bm\Theta )   \right>_{q^{(i)}(\bar\W) q^{(i)}(\bm\Gamma^*)  q^{(i)}(\bm\Gamma^v)  q^{(i)}(\Z)  }\label{eqre1}        \\
\propto&  \left<  \ln p(\Y|\bar\W, \alpha) \right>_{q^{(i)}(\bar\W)} +  \ln p(\alpha)    \notag\\
\propto& (a + KT -1)\ln \alpha \notag\\
 &-  \alpha\left( b+ \sum_{k=1}^K \left(\| \y_k - \bm\Phi \bm\mu_k^{(i)}    \|_2^2     + \tr(\bm\Phi \bm\Sigma_k^{(i)}\bm\Phi^H)\right)    \right),
\end{align}
where $\bm\mu_k^{(i)} \triangleq \left<  {\w}_k \right>_{q^{(i)}(\bar\w_k)} $ and $\bm\Sigma_k^{(i)} \triangleq  \left<  ({\w}_k - \bm\mu_k^{(i)} )({\w}_k - \bm\mu_k^{(i)} )^H\right>_{q^{(i)}(\bar\w_k)}$ [whose closed-form expressions are given in (\ref{eqclmu}) and (\ref{eqclSigma})]. Hence, $q^{(i+1)}(\alpha)$ obeys a gamma distribution:
\begin{align}
q^{(i+1)}(\alpha)=& \Gamma(\alpha|  a_\alpha^{(i)}, b_\alpha^{(i)}  ),
\end{align}
where $ a_\alpha^{(i)}= (a + KT ) $ and $b_\alpha^{(i)} = b+ \sum_{k=1}^K (\| \y_k - \bm\Phi \bm\mu_k^{(i)}    \|_2^2     + \tr(\bm\Phi \bm\Sigma_k^{(i)}\bm\Phi^H))$.

\subsection{Proof of Lemma~4}\label{lemmab}

Following a similar derivation to (\ref{eq-dq1})--(\ref{eq-dq16}), the optimization problem (\ref{eqM2}) has a unique solution:
\begin{align}
&\ln q^{(i+1)}(\bar\W)\notag\\
\propto&   \left<   \ln p( \Y, \bm\Theta )   \right>_{q^{(i+1)}(\alpha)q^{(i)}(\bm\Gamma^*)  q^{(i)}(\bm\Gamma^v)  q^{(i)}(\Z)  } \label{eqre2}\\
\propto&
\sum_{k=1}^K \left<   \ln p( \y_k, \bm\Theta )   \right>_{q^{(i+1)}(\alpha)q^{(i)}(\bm\Gamma^*)  q^{(i)}(\bm\gamma^v_k)  q^{(i)}(\Z)  }.\label{eq-upw1}
\end{align}
For each term in (\ref{eq-upw1}), we have
\begin{align}
&\left<   \ln p( \y_k, \bm\Theta )   \right>_{q^{(i+1)}(\alpha)q^{(i)}(\bm\Gamma^*)  q^{(i)}(\bm\gamma^v_k)  q^{(i)}(\z_k)  }\notag\\
\propto&\left<   \ln p( \y_k | \bar\w_k )           \right>_{q^{(i+1)}(\alpha)}
+ \left< \ln   p(\w_k^v| \bm{\gamma}^v_k)      \right>_{ q^{(i)}(\bm\gamma^v_k)}\notag\\
  &~~~~~~~~~~~~~~~~~~~~ +  \left<  \ln p(\w_k^s | \z_k, \bm\Gamma^* )   \right>_{q^{(i)}(\bm\Gamma^*) q^{(i)}(\z_k)  }\\
\propto&   -\hat{\alpha}^{(i+1)}\| \y_k- \bar{\bm\Phi}\bar{\w}_k      \|_2^2
- \rho^{-1}  (\w_k^v)^H \diag\left((\hat{\bm\gamma}^{v}_k)^{(i)}\right)\w_k^v\notag\\
& ~~~~-  (\w_k^s)^H \diag\underbrace{\left(\sum_{g=1}^G   \hat\phi^{(i)}_{k,g}(\hat{\bm\gamma}^{*}_g)^{(i)}\right)}_{\triangleq(\hat{\bm\gamma}^{s}_k)^{(i)}}\w_k^s,
\end{align}
where $(\hat{\bm\gamma}^{v}_k)^{(i)}= \left< {\bm\gamma}^{v}_k \right>_{ q^{(i)}(\bm\gamma^v_k)}$
and $(\hat{\bm\gamma}^{*}_g)^{(i)}= \left< {\bm\gamma}^{*}_g \right>_{ q^{(i)}(\bm\gamma^*_g)}$.

This equality shows that $q^{(i+1)}(\bar\W)$ is separable for each $\bar\w_k$, and $q^{(i+1)}(\bar\w_k)$ follows a Gaussian distribution:
\begin{align}
q^{(i+1)}(\bar\w_k)=  \mathcal{CN}(\bar\w_k| \bar{\bm\mu}_k^{(i+1)} ,  \bar{\bm\Sigma}_k^{(i+1)}   ),
\end{align}
where $\bar{\bm\mu}_k^{(i+1)}= \hat{\alpha}^{(i+1)} \bar{\bm\Sigma}_k^{(i+1)} \bar{\bm\Phi}^H  \y_k  $ and $\bar{\bm\Sigma}_k^{(i+1)}  =  \left(\hat{\alpha}^{(i+1)} \bar{\bm\Phi}^H \bar{\bm\Phi}  + \diag\left([(\hat{\bm\gamma}^{s}_k)^{(i)};
\rho^{-1}(\hat{\bm\gamma}^{v}_k)^{(i)}]  \right)\right)^{-1}$.

\subsection{Proof of Lemma~5}\label{lemmac}

Following a similar derivation to (\ref{eq-dq1})--(\ref{eq-dq16}), the optimization problem (\ref{eqM3}) has a unique solution:
\begin{align}
&\ln q^{(i+1)}(\bm\Gamma^*)\notag\\
\propto&   \left<\ln p( \Y, \bm\Theta )   \right>_{q^{(i+1)}(\alpha)q^{(i+1)}(\bar\W)  q^{(i)}(\bm\Gamma^v)  q^{(i)}(\Z)  } \label{eqre3}\\
\propto&  \sum_{k=1}^K \left<\ln p(\w_k^s | \z_k, \bm\Gamma^* )   \right>_{q^{(i+1)}(\bar\W) q^{(i)}(\Z)  }
+\sum_{g=1}^G\ln p(\bm\gamma_g^*) \label{eq-upgs1}\\
\propto& -\sum_{g=1}^G  \sum_{k=1}^K
\sum_{l=1}^{\hat L}\hat\phi_{k,g}^{(i)} \gamma^*_{g,l}\left<(w_{k,l}^s)^* w_{k,l}^s \right>_{ q^{(i+1)}(\bar\w_k)  }\notag\\
& + \sum_{g=1}^G  \sum_{k=1}^K
\sum_{l=1}^{\hat L}\hat\phi_{k,g}^{(i)} \ln \gamma^*_{g,l} + \sum_{g=1}^G\sum_{l=1}^{\hat L} ( (a-1) \ln \gamma^*_{g,l}- b  \gamma^*_{g,l}).
\end{align}

Clearly, $q^{(i+1)}(\bm\Gamma^*)$ is separable for each $\gamma^*_{g,l}$, and we obtain
\begin{align}
\ln q^{(i)}(\gamma^*_{g,l})\propto& - \gamma^*_{g,l} \left( b +  \sum_{k=1}^K  \hat\phi_{k,g}^{(i)} \left( |\bar{\mu}_{k,1,l}^{(i+1)}|^2 +\bar{\Sigma}_{k,1,l}^{(i+1)}  \right)     \right)\notag\\
&~~+  \left( a -1 +  \sum_{k=1}^K\hat\phi_{k,g}^{(i)}\right)\ln \gamma^*_{g,l},
\end{align}
where $\hat\phi_{k,g}^{(i)}\triangleq q^{(i)} (z_{k,g}=1)$, $\bar{\mu}_{k,1,l}^{(i+1)}$ stands for the $l$-th element of $\bar{\bm\mu}_{k,1}^{(i+1)}$,  and  $\bar{\Sigma}_{k,1,l}^{(i+1)}$ stands for the $l$-th diagonal element of $\bar{\bm\Sigma}_{k,1}^{(i+1)}$.
Hence, $q^{(i+1)}(\gamma^*_{g,l})$ obeys a gamma distribution:
\begin{align}
q^{(i+1)}(\gamma^*_{g,l})= \Gamma\left(\gamma^*_{g,l} | (a^*_{g,l})^{(i+1)}, (b^*_{g,l})^{(i+1)}\right)
\end{align}
with $(a^*_{g,l})^{(i+1)}= a +  \sum_{k=1}^K\hat\phi_{k,g}^{(i)}$ and $ (b^*_{g,l})^{(i+1)}= b +  \sum_{k=1}^K  \hat\phi_{k,g}^{(i)} ( |\bar{\mu}_{k,1,l}^{(i+1)}|^2 +\bar{\Sigma}_{k,1,l}^{(i+1)}  )$.

\subsection{Proof of Lemma~7}\label{lemmad}

Following a similar derivation to (\ref{eq-dq1})--(\ref{eq-dq16}), the optimization problem (\ref{eqM5}) has a unique solution:
\begin{align}
&\ln q^{(i+1)}(\Z)\notag\\
\propto&   \left<\ln p( \Y, \bm\Theta )   \right>_{q^{(i+1)}(\alpha)q^{(i+1)}(\bar\W) q^{(i+1)}(\bm\Gamma^*) q^{(i+1)}(\bm\Gamma^v) }\label{eqre4}\\
\propto&  \sum_{k=1}^K \left<\ln p(\w_k^s | \z_k, \bm\Gamma^* )   \right>_{q^{(i+1)}(\bar\W) q^{(i+1)}(\bm\Gamma^*)  }. \label{eq-up5z}
\end{align}
From (\ref{eq-up5z}) and the fact that $\z_k$ is a discrete vector, we are able to exhaustively calculate the value of $\ln q^{(i+1)} (z_{k,g}=1) $, $\forall k,g$ as
\begin{align}
&\ln q^{(i+1)} (z_{k,g}=1)\notag\\
\propto& \underbrace{\sum_{l=1}^{\hat L} (\widehat{\ln \gamma^*_{g,l}})^{(i+1)} -
\sum_{l=1}^{\hat L} (\gamma^*_{g,l})^{(i+1)}\left( |\bar{\mu}_{k,1,l}^{(i+1)}|^2 +\bar{\Sigma}_{k,1,l}^{(i+1)}  \right)}_{= \varsigma_{k,g}^{(i+1)}}\notag.
\end{align}
Since $ \sum_{g=1}^{G} q^{(i+1)} (z_{k,g}=1)=1$, we obtain
\begin{align}
\hat\phi_{k,g}^{(i+1)}= q^{(i+1)} (z_{k,g}=1)  =  \frac{\exp(\varsigma_{k,g}^{(i+1)})}{\sum_{g=1}^G \exp( \varsigma_{k,g}^{(i+1)}  )}.
\end{align}

\subsection{Proof of Lemma~8}\label{lemmae}
The non-decreasing property can be achieved by
\begin{align}
&\mathcal{U}(  q_{1}^{(i)}, q_{2}^{(i)},q_{3}^{(i)},q_{4}^{(i)},q_{5}^{(i)}   )\notag\\
\le& \mathcal{U}(  q_{1}^{(i+1)}, q_{2}^{(i)},q_{3}^{(i)},q_{4}^{(i)},q_{5}^{(i)}   )\label{eqal1}\\
\le& \mathcal{U}(  q_{1}^{(i+1)}, q_{2}^{(i+1)},q_{3}^{(i)},q_{4}^{(i)},q_{5}^{(i)}   )\label{eqal2}\\
\le& \mathcal{U}(  q_{1}^{(i+1)}, q_{2}^{(i+1)},q_{3}^{(i+1)},q_{4}^{(i)},q_{5}^{(i)}   )\label{eqal3}\\
\le& \mathcal{U}(  q_{1}^{(i+1)}, q_{2}^{(i+1)},q_{3}^{(i+1)},q_{4}^{(i+1)},q_{5}^{(i)}   )\label{eqal4}\\
\le& \mathcal{U}(  q_{1}^{(i+1)}, q_{2}^{(i+1)},q_{3}^{(i+1)},q_{4}^{(i+1)},q_{5}^{(i+1)}   ),\label{eqal5}
\end{align}
where (\ref{eqal1}), (\ref{eqal2}), (\ref{eqal3}), (\ref{eqal4}) and (\ref{eqal5}) follow (\ref{eqM1}), (\ref{eqM2}), (\ref{eqM3}), (\ref{eqM4}) and (\ref{eqM5}), respectively.

\subsection{Proof of Lemma 9}\label{lemmaStationary}
From Section III-C,  it is clear that $q(\bm\Theta )=q(\alpha) q(\bar\W) q(\bm\Gamma^*) q(\bm\Gamma^v)q(\Z)$ can be considered as some parameterized functions,  e.g., a gamma distribution function with parameters $a_\alpha$ and $b_\alpha$ for $q(\alpha)$, a Gaussian distribution function with parameters $ \bar{\bm\mu}_k$s and $\bar{\bm\Sigma}_k$s for $q(\bar\W)$, and so on.
As a result, the optimization problem (\ref{eq-all1}) which is optimized over function spaces can be converted into a conventional parameter optimization problem. Therefore, the definition and convergence result for the conventional stationary point can be applied.

Let the surrogate function be chosen as the objective function itself, and then,
according to Theorem 2-b in \cite{razaviyayn2014successive}, the proposed algorithm converges to a stationary solution  because the problems in (\ref{eqM1})--(\ref{eqM5}) have a unique solution.

\subsection{Derivation for Eq. (\ref{eqderbeta})}\label{lemmaoff}

Ignoring the independent terms, the objective function in (\ref{eq2M6}) becomes
\begin{align}
& \mathcal{U}(  q_{1}^{(i+1)}, q_{2}^{(i+1)},q_{3}^{(i+1)},q_{4}^{(i+1)},q^{(i+1)}_{5},\B    )\notag\\
\propto&   \left<\sum_{k=1}^K \ln p(\y_k | \w_k^s, \w_k^v, \alpha,\bm\beta_k,\bm\varphi_k)  \right>_{q^{(i+1)}(\alpha)q^{(i+1)}(\bar\W)}\notag\\
\propto&  -\hat\alpha^{(i+1)}\sum_{k=1}^K \left\| \y  -  \bm\Phi(\bm\beta_k,\bm\varphi_k) \bm\mu_k^{(i+1)} \right\|_2^2\notag\\
&- \hat\alpha^{(i+1)} \sum_{k=1}^K \tr\left(\bm\Phi(\bm\beta_k,\bm\varphi_k)\bm\Sigma_k^{(i+1)}\bm\Phi^H(\bm\beta_k,\bm\varphi_k)\right). \notag
\end{align}
Obviously, the objective function is separable for each $\bm\beta_k$.  Calculating the derivative of each term in  the above equality  w.r.t.  $\beta_{k,l}$, we obtain
\begin{align*}
&\frac{\partial \left\|\y- \bm\Phi(\bm\beta_k,\bm\varphi_k) \bm\mu_k^{(i+1)} \right\|_2^2}{\partial \beta_{k,l} }\\
=&\frac{\partial\left\|\y^{(i+1)}_{k-l} - \mu_{k,l}^{(i+1)} \cdot\X (\a (\hat\vartheta_{ l}+\beta_{k,l},\varphi_{k,l}) )  \right\|_2^2}{\partial \beta_{k,l} } \notag\\
=& 2 \mathrm{Re}\left( (  \a' (\hat\vartheta_{ l}+\beta_{k,l},\varphi_{k,l}))^H \X^H\X \a (\hat\vartheta_{ l}+ \beta_{k,l},\varphi_{k,l})\right) \cdot  |\mu_{k,l}^{(i+1)}|^2\notag\\
 &~~~~~~~~   - 2 \mathrm{Re}\left(  (\a' (\hat\vartheta_{ l}+ \beta_{k,l},\varphi_{k,l}))^H \X^H \cdot (\mu_{k,l}^{(i+1)})^* \y_{k-l}^{(i+1)}\right)
\end{align*}
and
\begin{align*}
&\frac{\partial \tr\left(\bm\Phi(\bm\beta_k,\bm\varphi_k)\bm\Sigma_k^{(i+1)}\bm\Phi^H(\bm\beta_k,\bm\varphi_k)\right) }{\partial \beta_{k,l} }\notag\\
=&2 \mathrm{Re}\left(   (\a' (\hat\vartheta_{ l}+\beta_{k,l},\varphi_{k,l}))^H \X^H\X \a (\hat\vartheta_{ l}+ \beta_{k,l},\varphi_{k,l}) \right) \cdot \chi_{k,ll}^{(i+1)}\notag\\
+& 2 \mathrm{Re}\left(  (\a' (\hat\vartheta_{ l}+ \beta_{k,l},\varphi_{k,l}))^H \X^H \X \sum_{j\ne l} \chi_{k,jl}^{(i+1)} \a (\hat\vartheta_{j}+ \beta_{k,j}, \varphi_{k,j})\right),
\end{align*}
where $\y_{k-l}^{(i+1)}= \y_k -   \X \cdot\sum_{j\ne l } (\mu_{k,j}^{(i+1)} \cdot\a (\hat\vartheta_{j}+ \beta_{k,j},\varphi_{k,l})) $, $\a' (\hat\vartheta_{l}+\beta_{k,l},\varphi_{k,l})= d \a(\hat\vartheta_{l}+\beta_{k,l},\varphi_{k,l}) /{d \beta_{k,l}} $, and $\mu_{k,l}^{(i+1)}$ and $\chi_{k,jl}^{(i+1)}$ denote the $l$-th element and the $(j,l)$-th element of $\bm\mu_{k}^{(i+1)}$ and $\bm\Sigma_{k}^{(i+1)}$, respectively.
Hence, the derivative element $\zeta^{(i+1)}(\beta_{k,l})$ in (\ref{eqderbeta1}) is achieved.


\bibliographystyle{IEEEtran}
\bibliography{groupCE}

\end{document}